%% file: arxiv.tex
\documentclass[10pt,twocolumn,letterpaper]{article}

\usepackage[pagenumbers]{cvpr} % To force page numbers, e.g. for an arXiv version

\input{preamble}
\definecolor{cvprblue}{rgb}{0.21,0.49,0.74}
\usepackage[pagebackref,breaklinks,colorlinks,allcolors=cvprblue]{hyperref}

\usepackage[table]{xcolor}  % table 选项很关键
\usepackage{colortbl}       % 支持 \cellcolor
\usepackage{wrapfig}
\usepackage{url}
\usepackage{amsthm}
\usepackage{graphicx}
\usepackage{booktabs}
\usepackage{listings}
\usepackage{makecell}   % 表头换行
\usepackage{siunitx}    % 数值对齐
\usepackage{multirow}
\usepackage{balance}

\usepackage{silence}
\def\paperID{23569} % *** Enter the Paper ID here
\def\confName{CVPR}
\def\confYear{2026}

\newcommand{\graycell}[1]{\textcolor{gray}{\textit{#1}}}
\newcommand{\bluecell}[1]{\textcolor{cyan}{\textit{#1}}}
\newcommand{\redcell}[1]{\textcolor{red}{\textit{#1}}}

\title{DiT-IC: Aligned Diffusion Transformer for Efficient Image Compression}

\author{
Junqi Shi, Ming Lu, Xingchen Li, Anle Ke, Ruiqi Zhang, Zhan Ma\\
School of Electronic Science and Engineering, Nanjing University, Nanjing 210023, China\\
Corresponding author: {\tt \{minglu, mazhan\}@nju.edu.cn} \\
Code: \url{https://njuvision.github.io/DiT-IC/}
}

\begin{document}
% \maketitle

\twocolumn[{
\maketitle
% \vspace{-1em}
\begin{center}
    \includegraphics[width=0.97\linewidth]{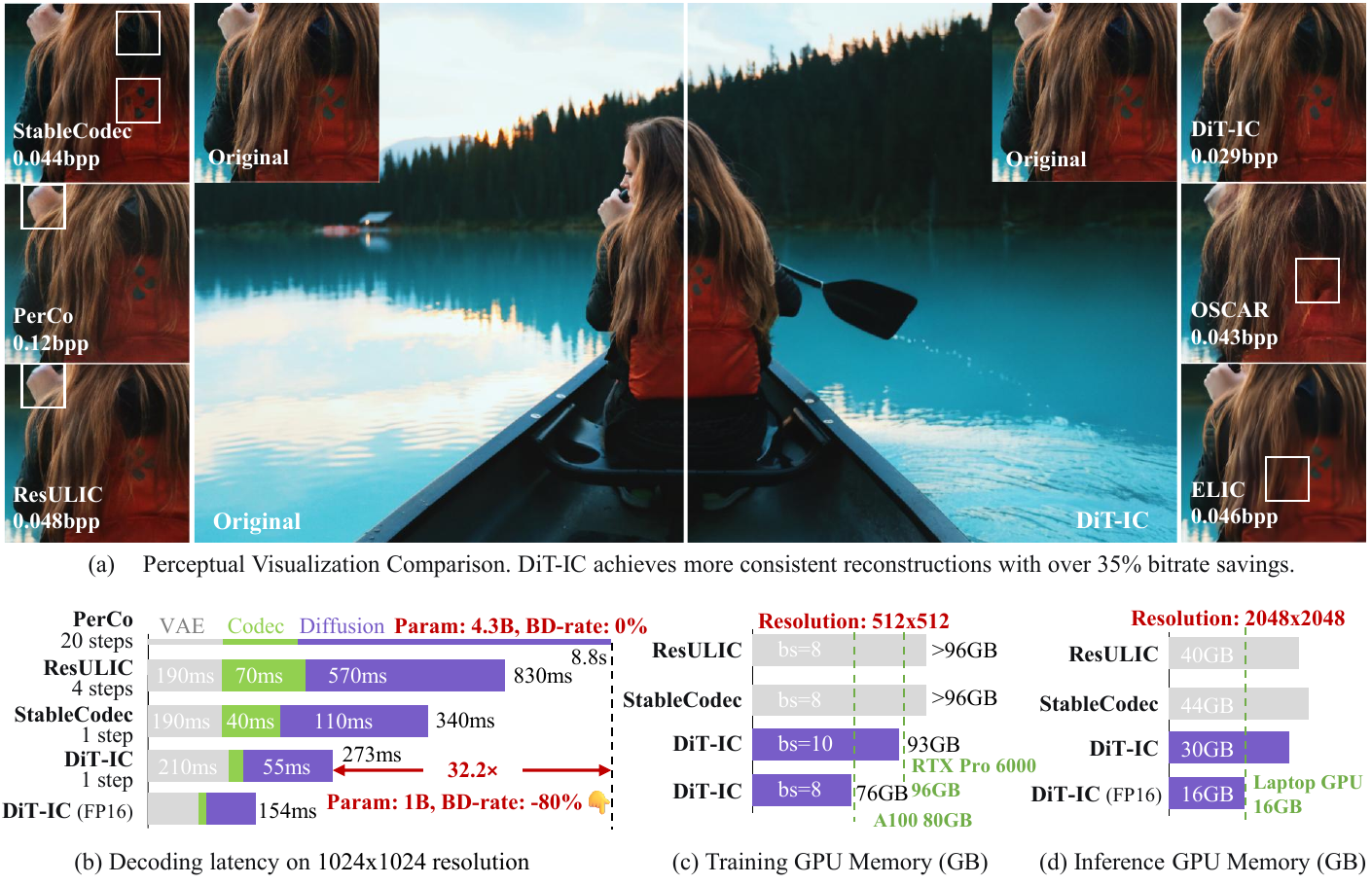}
    % \vspace{-5pt}
    \phantomsection
    \captionof{figure}{
        \textbf{Overview of reconstructed results and efficiency of our proposed DiT-IC.}
    }
    \label{fig:head}
    % \vspace{10pt}
\end{center}
}]

\input{sec/0_abstract}    
\input{sec/1_intro}

\begin{figure*}[ht]
\centering
\includegraphics[width=0.95\textwidth]{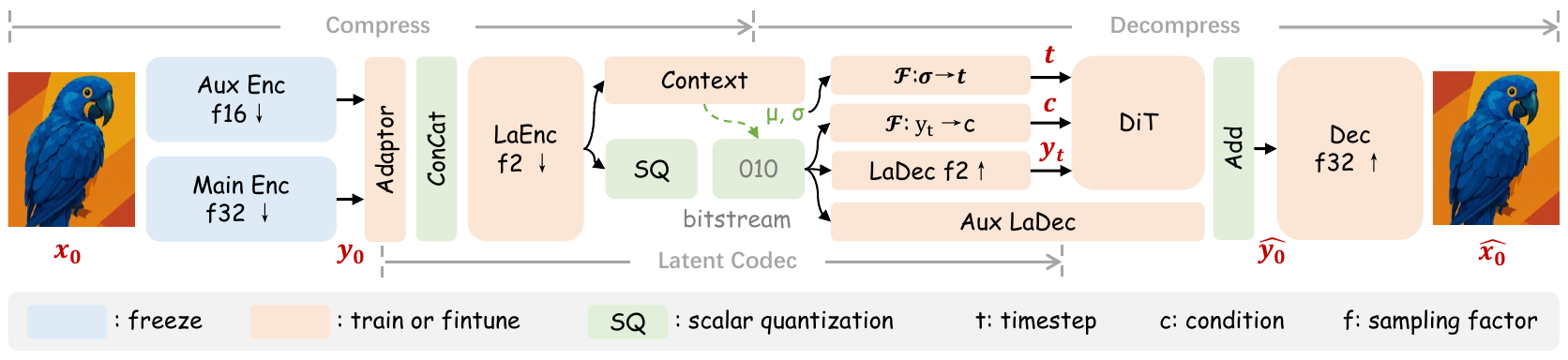}
\caption{\textbf{Overview of the proposed DiT-IC framework.} Following StableCodec~\cite{zhang2025stablecodec}, we adopt ELIC~\cite{he2022elic} as our auxiliary encoder.}
\label{fig:arch}
\end{figure*}

\input{sec/2_preliminary}

\input{sec/3_methods}

\input{sec/4_experiment}

\input{sec/5_conclusion}

\clearpage
\newpage
\input{sec/X_suppl}

\clearpage
\section*{Acknowledgments}
This work was supported in part by Natural Science Foundation of China
(Grant No. 62401251, 62431011) and Natural Science
Foundation of Jiangsu Province (Grant No. BK20241226, BK20243038). The
authors would like to express their sincere gratitude to the
Interdisciplinary Research Center for Future Intelligent Chips
(Chip-X) and Yachen Foundation for their invaluable support.

{
    \small
    \balance
    \bibliographystyle{ieeenat_fullname}
    \bibliography{main}
}

\end{document}

%% file: sec/0_abstract.tex
\begin{abstract}
Diffusion-based image compression has recently shown outstanding perceptual fidelity, yet its practicality is hindered by prohibitive sampling overhead and high memory usage.
Most existing diffusion codecs employ U-Net architectures, where hierarchical downsampling forces diffusion to operate in shallow latent spaces (typically with only $8\times$ spatial downscaling), resulting in excessive computation.
In contrast, conventional VAE-based codecs work in much deeper latent domains ($16\times$–$64\times$ downscaled), motivating a key question:
\emph{Can diffusion operate effectively in such compact latent spaces without compromising reconstruction quality?}
To address this, we introduce \textbf{DiT-IC}—an Aligned \textbf{Di}ffusion \textbf{T}ransformer for \textbf{I}mage \textbf{C}ompression—which replaces the U-Net with a Diffusion Transformer capable of performing diffusion in latent space entirely at $32\times$ downscaled resolution.
DiT-IC adapts a pretrained text-to-image multi-step DiT into a single-step reconstruction model through three key alignment mechanisms:
(1) a \textbf{variance-guided reconstruction flow} that adapts denoising strength to latent uncertainty for efficient reconstruction;
(2) a \textbf{self-distillation alignment} that enforces consistency with encoder-defined latent geometry to enable one-step diffusion; and
(3) a \textbf{latent-conditioned guidance} that replaces text prompts with semantically aligned latent conditions, enabling text-free inference.
With these designs, DiT-IC achieves state-of-the-art perceptual quality while offering up to \textbf{30× faster decoding} and drastically lower memory usage than existing diffusion-based codecs. Remarkably, it can reconstruct $2048\times2048$ images on a 16 GB laptop GPU.
\end{abstract}

%% file: sec/1_intro.tex
\section{Introduction}
\label{sec:intro}
Recent diffusion-based generative models~\cite{rombach2022high, esser2024scaling, yao2025reconstruction} have achieved remarkable advances in visual synthesis, producing photorealistic and semantically controllable images. However, for a fundamental low-level task like image compression, which demands practical efficiency—low latency and memory economy—most diffusion-based compression approaches~\cite{careil2023towards, keultra} remain constrained by heavy sampling overhead and substantial memory usage.

A key source of inefficiency lies in the spatial scale where diffusion operates. Existing diffusion-based codecs typically perform denoising in relatively shallow latent spaces (e.g., $8\times$ spatial reduction), resulting in significant computational and memory burdens. In contrast, modern learned codecs naturally operate in much deeper latent domains, often with $16\times$~\cite{he2022elic, liu2023learned}, $32\times$~\cite{keultra}, or even $64\times$~\cite{duan2023qarv, zhang2025stablecodec} spatial reductions.
This discrepancy motivates a central question:
\textit{Can diffusion operate effectively in deeply compressed latent spaces to enable efficient reconstruction without sacrificing fidelity?}

Most existing diffusion-based codecs~\cite{keultra, zhang2025stablecodec, li2025rdeic, guo2025oscar} employ U-Net-based diffusion architectures, whose hierarchical downsampling further reduces the spatial scale (Fig.~\ref{fig:intro}), making them poorly suited for deeply compressed latents.
Recently, \emph{Diffusion Transformers (DiTs)}~\cite{peebles2023scalable, chen2023pixart, xie2024sana} have emerged as a compelling alternative, replacing U-Nets with cascaded transformer blocks that maintain a constant spatial resolution throughout the denoising process. This architectural property makes DiTs naturally compatible with deeply compressed latent domain and provides a promising foundation for efficient diffusion modeling.

However, directly transplanting pretrained diffusion models into compact, compression-oriented latent spaces often results in severe degradation. The core challenge lies in the mismatch between \emph{generative} and \emph{reconstructive} objectives.
Unlike text-to-image diffusion, which begins from pure Gaussian noise, image compression starts from structured, entropy-constrained latents that already lie near the data manifold. This structured initialization significantly narrows the sampling distribution, suggesting that iterative multi-step denoising may be redundant—and that even single-step reconstruction could be achievable.
Yet naïvely fine-tuning generative diffusers fails to exploit this property, often leading to misaligned feature manifolds and suboptimal reconstructions.

To overcome these challenges, we propose \textbf{DiT-IC}—an Aligned \textbf{Di}ffusion \textbf{T}ransformer for \textbf{I}mage \textbf{C}ompression—which adapts a pretrained text-to-image multi-step DiT into an efficient one-step reconstruction model operating in a $32\times$ latent diffusion space.
Our method introduces three complementary alignment mechanisms that jointly bridge the gap between diffusion generation and compression reconstruction:

\begin{figure}[t]
    \centering
    \includegraphics[width=0.95\columnwidth]{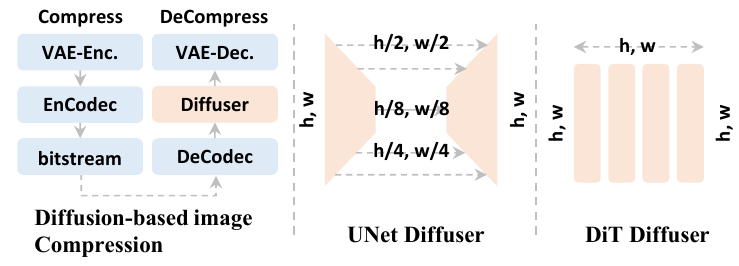}
    \caption{\textbf{Architectural comparison.} The left panel illustrates the overall diffusion-based image compression framework.
    U-Net-based diffusers perform multi-stage downsampling, while DiTs maintain a constant spatial resolution throughout the denoising process, making them naturally compatible with deeply compressed latent inputs.
    }
    \label{fig:intro}
\end{figure}

\textbf{Variance-Guided Reconstruction Flow.}
We reinterpret the denoising trajectory as an adaptive reconstruction flow, where spatially varying uncertainty determines local denoising strength. By mapping latent variance to pseudo-timesteps, DiT-IC collapses iterative denoising into a single transformation that preserves fine details while maintaining decoding efficiency.

\textbf{Self-Distillation Alignment.}
To stabilize one-step learning without external supervision, we introduce a self-distillation mechanism that enforces consistency between the denoised output and the encoder’s frozen latent representation, effectively distilling multi-step diffusion behavior into a single forward pass.

\textbf{Latent-Conditioned Guidance.}
We replace text-based conditioning with a lightweight latent-conditioned projection derived from compressed representations. By contrastively co-aligning latent and textual embeddings during training, the model retains semantic priors from the pretrained DiT while eliminating the need for text input during inference, thereby removing the heavy text encoder.

With these designs, \textbf{DiT-IC} achieves state-of-the-art rate–distortion performance, offering up to \textbf{30× faster decoding} and substantially lower memory usage than existing diffusion-based codecs.
Remarkably, it can reconstruct $2048\times2048$ images on a 16 GB laptop GPU.
Our findings demonstrate that pretrained diffusion transformers, when properly aligned with compression objectives, can serve as powerful one-step reconstruction priors for efficient visual compression.

%% file: sec/2_preliminary.tex
\section{Preliminary}
\label{sec:preliminary}

\textbf{Diffusion Transformers.}
Diffusion models~\cite{ho2020denoising, song2020score} synthesize data by iteratively denoising Gaussian noise toward the data manifold.
Early variants typically adopt U-Net backbones~\cite{rombach2022high}, whose multi-scale encoder–decoder structures provide strong spatial locality but suffer from limited scalability and global consistency.
\emph{Diffusion Transformers (DiTs)}~\cite{peebles2023scalable} overcome these issues by replacing the U-Net with cascaded transformer blocks that operate at a \emph{single, constant spatial resolution}.
This design eliminates hierarchical downsampling, enabling globally coherent representation learning and improved scalability.
Recent extensions—such as PixArt-$\alpha$~\cite{chen2023pixart}, SD3~\cite{esser2024scaling}, and Flux~\cite{flux2024}—have scaled DiTs to large, multimodal generation, while efficient variants like Sana~\cite{xie2024sana} accelerate inference via linear attention mechanisms.

\textbf{Flow Matching.}
Traditional diffusion models can be interpreted as discretized stochastic differential equations (SDEs) that gradually transform noise into data through a stochastic denoising process.
\emph{Flow Matching (FM)}~\cite{lipman2022flow, liu2022flow} reformulates this paradigm as a deterministic ordinary differential equation (ODE), where a neural network learns a velocity field that continuously transports samples from a simple prior to the data manifold~\cite{tong2023improving}.
Compared with stochastic diffusion, FM provides a mathematically elegant and computationally efficient framework, often enabling faster sampling and consistent results.

\textbf{Diffusion-based Image Compression.}
Recent studies have explored integrating diffusion models into learned image compression (LIC) to leverage their powerful generative priors for perceptually faithful reconstruction.
Early methods such as DiffEIC~\cite{pan2022extreme}, \citet{yang2023lossy}, and CDC~\cite{li2024towards} encode images into compact latent conditions that guide pretrained diffusion models during reconstruction.
Later approaches—including RDEIC~\cite{li2025rdeic} and ResULIC~\cite{keultra}—reinterpret the denoising trajectory as a progressive reconstruction process, where each diffusion step refines the compressed representation.
Other works~\cite{relic2024lossy, relic2025bridging} employ diffusion priors for post-quantization enhancement.
Recent advances—such as StableCodec~\cite{zhang2025stablecodec}, OneDC~\cite{xue2025one}, and OSCAR~\cite{guo2025oscar}—further improve efficiency by collapsing multi-step diffusion into single-step inference. Existing methods often adapts pretrained Stable Diffusion models~\cite{rombach2022high, podell2023sdxl} for practical codec deployment.

%% file: sec/3_methods.tex
\section{Method}
\label{sec:method}
We introduce DiT-IC, an aligned Diffusion Transformer framework for efficient image compression.
Unlike U-Net-based diffusion models that operate in shallow latent spaces (typically $8\times$), DiT-IC performs diffusion directly within a deeper $32\times$ latent domain, achieving higher efficiency while maintaining perceptual fidelity.
To adapt the generative diffusion process to the reconstruction-oriented objective of compression, DiT-IC incorporates a set of alignment mechanisms across three key dimensions:
(1) from generation to reconstruction, aligning diffusion strength with latent variance;
(2) from multi-step to single-step inference, improving efficiency without compromising quality; and
(3) from text-guided to latent-conditioned diffusion, enabling text-free decoding.
Together, these components form an end-to-end aligned diffusion framework that delivers efficient image compression, as illustrated in Fig.~\ref{fig:arch}.

\subsection{Variance-Guided Reconstruction Flow: From Generation to Reconstruction}
\begin{figure}[t]
    \centering
    \includegraphics[width=0.95\columnwidth]{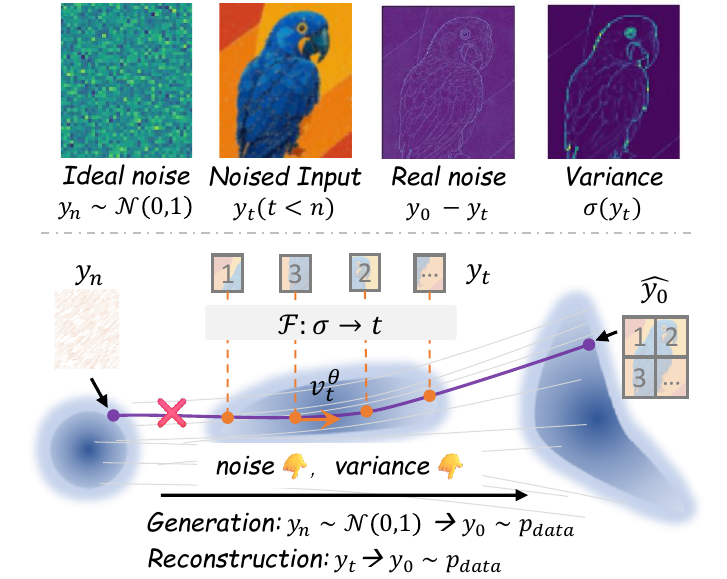}
    \caption{\textbf{Variance-Guided Flow Matching.}
    Unlike standard diffusion that starts from Gaussian noise, compression reconstruction begins from a quantized latent ${\mathbf{y}}_t$ containing structured noise.
    The local variance $\sigma(\mathbf{y}_t)$ measures spatial uncertainty, which we map to pseudo-timesteps $t = \mathcal{F}(\sigma)$ for spatially adaptive one-step flow matching.}
    \label{fig:flow1}
\end{figure}

Traditional flow matching~\cite{lipman2022flow} learns a continuous vector field to transport samples from Gaussian noise $\mathcal{N}(0, I)$ to the data distribution $p_\text{data}$ through the probability flow ODE:
\begin{equation}
\frac{d\mathbf{y}_t}{dt} = \mathbf{v}_\theta(\mathbf{y}_t, t), \quad \mathbf{y}_T \sim \mathcal{N}(0, I).
\label{eq:pf_ode}
\end{equation}
However, in image compression, the initial state is not pure noise but a quantized latent $\mathbf{y}_t$ that already lies close to the data manifold.
This observation motivates a \textbf{one-step reconstruction flow}, replacing the iterative denoising process with a single adaptive transformation.

As shown in Fig.~\ref{fig:flow1}, compression noise $\mathbf{y}_0 - \mathbf{y}_t$ exhibits strong spatial heterogeneity—smooth areas behave like low-noise (small-timestep) regions, while textured regions resemble high-noise (large-timestep) states.
Therefore, a single global timestep cannot adequately model the local noise characteristics.

To address this, we introduce a variance-guided pseudo-timestep mapping.
In learned compression, the latent distribution is typically parameterized as $\mathbf{y} \sim \mathcal{N}(\boldsymbol{\mu}, \boldsymbol{\sigma}^2)$, where the predicted mean $\boldsymbol{\mu}$ captures the underlying image content, and variance $\boldsymbol{\sigma}$ reflects the uncertainty within that content.
We leverage this inherent uncertainty to define a differentiable mapping:
\begin{equation}
t = \mathcal{F}(\text{proj}_\theta(\boldsymbol{\sigma})) \in \mathbb{R}^{H \times W},
\label{eq:variance2time}
\end{equation}
where $\text{proj}_\theta(\cdot)$ projects $\boldsymbol{\sigma}$ to the latent dimension, and monotonic function $\mathcal{F}$ converts it to pixel-wise pseudo-timesteps.
Higher variance corresponds to larger $t$, indicating stronger denoising strength.

\begin{figure}[ht]
    \centering
    \includegraphics[width=0.8\columnwidth]{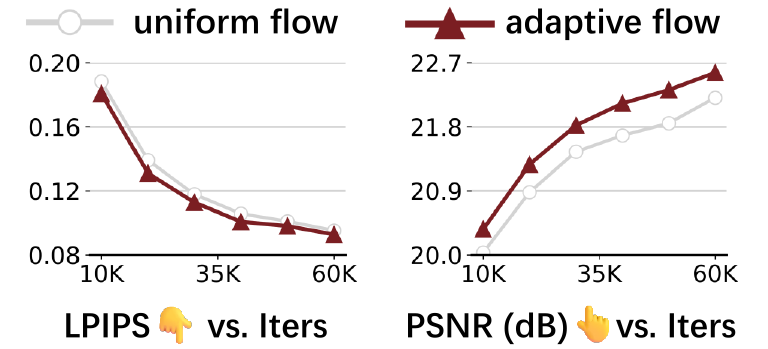}
    \caption{Ablation study of variance-guided reconstruction flow.}
    \label{fig:flow2}
\end{figure}

Given the adaptive timestep field $t = \mathcal{F}(\boldsymbol{\sigma})$, the one-step reconstruction is computed as:
\begin{equation}
\hat{\mathbf{y}} = \tilde{\mathbf{y}} - \mathbf{v}_\theta(\tilde{\mathbf{y}}, t),
\label{eq:adaptive_flow}
\end{equation}
which effectively collapses the multi-step denoising trajectory into a single spatially adaptive transformation—achieving high-fidelity reconstruction (Fig.~\ref{fig:flow2}).

Recent advances also explore adaptive generation strength.
UPSR~\cite{zhang2025uncertainty} leverages reconstruction error to adjust noise strength;
EAR~\cite{ma2025towards} modulates generation using image entropy;
and OSCAR~\cite{guo2025oscar} introduces a image-wise rate–timestep mapping for variable-rate diffusion.
Our method extends this idea to a \textbf{pixel-wise variance–timestep mapping}, providing finer-grained adaptation for one-step reconstruction.

\subsection{Self-Distillation Alignment: From Multi-Step to One-Step}
\begin{figure}[t]
\centering
\includegraphics[width=0.9\columnwidth]{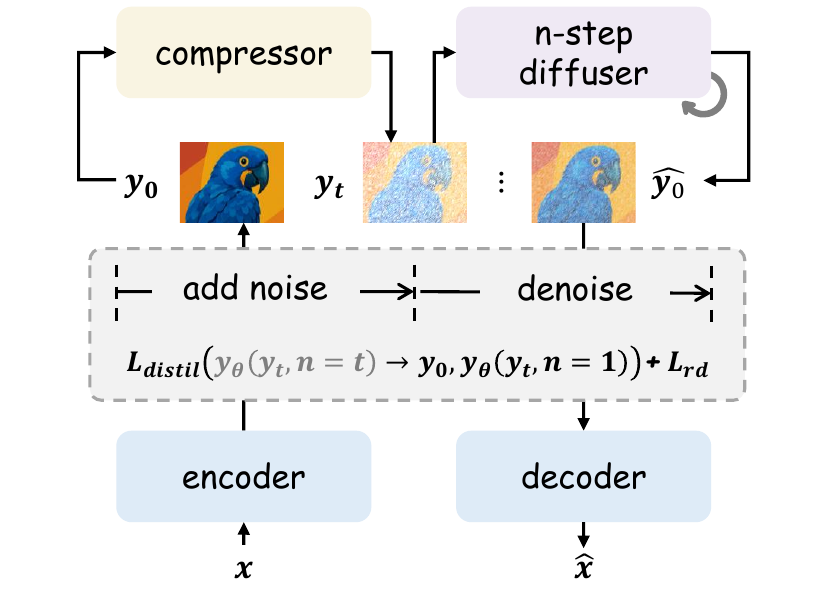}
\caption{\textbf{Self-Distillation Alignment.}
    DiT-IC distills the multi-step diffusion process into a single forward pass by aligning its denoised latent with the frozen encoder representation, while jointly optimizing the diffusion transformer and decoder.
    $\mathbf{y}_\theta(\cdot)$ denotes the denoised output, and $\mathcal{L}_{\text{rd}}$ indicates the rate–distortion loss.}
\label{fig:dist1}
\end{figure}
While the variance-guided flow enables adaptive one-step reconstruction, fine-tuning pretrained multi-step models remains challenging in the absence of explicit denoising trajectory supervision.
Conventional diffusion distillation frameworks~\cite{salimans2022progressive, meng2023distillation} depend on a pre-trained teacher to provide intermediate denoising trajectories—an approach infeasible for compression, where no multi-step reference exists within the deep latent domain.

To overcome this, we propose a \textbf{self-distillation alignment} strategy that replaces external supervision with an internal reference from the encoder.
The encoder’s latent output $\mathbf{y}_0$, which already lies close to the data manifold, naturally serves as a self-supervised target for the denoised latent $\hat{\mathbf{y}}_0$ predicted by the diffusion transformer.
We freeze the encoder and jointly optimize the DiT and decoder so that $\hat{\mathbf{y}}_0$ aligns with $\mathbf{y}_0$, effectively collapsing the multi-step denoising process into a deterministic, single-step reconstruction—while preserving the latent geometry defined by the encoder. The pipeline is shown in Fig.~\ref{fig:dist1}.

Formally, we apply a marginal cosine alignment loss:
\begin{equation}
\mathcal{L}_{\text{distil}} =
\mathbb{E}_{x \sim p_\text{data}}
\left[1 - m - \frac{\langle \hat{\mathbf{y}}, \mathbf{y}_0 \rangle}{|\hat{\mathbf{y}}|_2 |\mathbf{y}_0|_2}\right],
\label{eq:distil_components}
\end{equation}
where $m$ is a small margin encouraging angular separation between distinct latent directions.

\begin{figure}[ht]
    \centering
    \includegraphics[width=0.8\columnwidth]{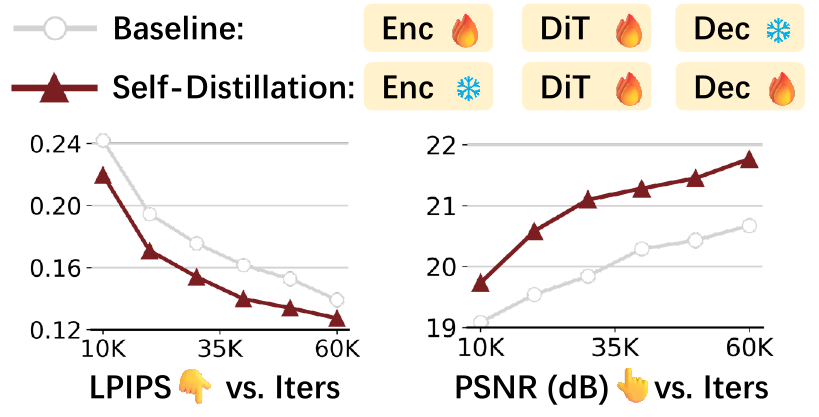}
    \caption{Ablation of self-distillation alignment.}
    \label{fig:dist2}
\end{figure}

In contrast to previous diffusion-based codecs~\cite{zhang2025stablecodec, xue2025one}, which partially fine-tune the encoder or freeze the decoder, our approach fixes the encoder while jointly adapting the DiT and decoder.
This co-adaptation stabilizes training, enhances perceptual fidelity, and supports efficient reconstruction in deep latent spaces, as illustrated in Fig.~\ref{fig:dist2}.

Conceptually, our self-distillation resembles feature-alignment paradigms used in generative modeling—such as VA-VAE~\cite{yao2025reconstruction} aligned with DINOv2~\cite{oquab2023dinov2}—but differs in that it leverages the frozen VAE encoder itself as an intrinsic alignment target for reconstruction.
Moreover, integrating recent adversarial distillation techniques~\cite{sauer2024adversarial} could further enhance the perceptual realism of reconstructed images in future work.

\subsection{Latent-Conditioned Guidance: From Text to Semantic Latent Condition}
\begin{figure}[t]
\centering
\includegraphics[width=0.9\columnwidth]{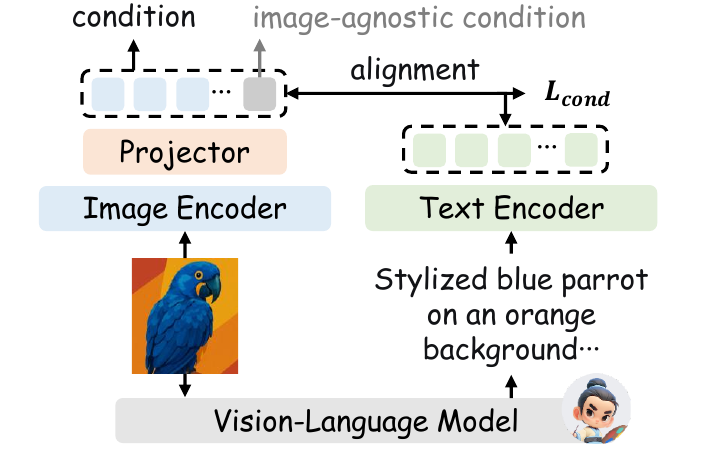}
\caption{\textbf{Latent-Conditioned Guidance.}
We replace text-based guidance in DiT with a latent-conditioned projection derived from the compressed representation by aligning projected latent and text embeddings, enabling text-free conditioning at inference.}
\label{fig:cond1}
\end{figure}

Pretrained diffusion transformers typically rely on text-conditioned guidance to control semantics. However, for reconstruction-oriented tasks, textual prompts are often inefficient and suboptimal: they may fail to capture fine-grained spatial structures and require large vision–language models (VLMs), introducing additional latency and stochasticity during inference.

We observe that the latent representation $\hat{y}$ itself encodes rich semantic and structural information, which can serve as an effective conditioning source. Motivated by this, we propose \textbf{Latent-Conditioned Guidance}, shown in Fig.~\ref{fig:cond1}, replacing the text condition $c_\text{text}$ with a learned latent condition:
\begin{equation}
c_\text{lat} = \text{Proj}_\psi(\hat{y}),
\label{eq:lcg_condition}
\end{equation}
where $\text{Proj}_\psi(\cdot)$ is a lightweight projection module mapping latent features into the same embedding space used by the pretrained text encoder in DiT.

To ensure semantic alignment, we perform \emph{contrastive co-alignment} between projected latent and text embeddings using a CLIP-style objective~\cite{radford2021learning}:
\begin{equation}
\mathcal{L}_{\text{cond}} =
-\mathbb{E}_{(x_i, t_i)} 
\left[
\log
\frac{\exp(\langle c_{\text{lat}, i}, c_{\text{text}, i} \rangle / \tau)}
{\sum_{j} \exp(\langle c_{\text{lat}, i}, c_{\text{text}, j} \rangle / \tau)}
\right],
\label{eq:align}
\end{equation}
where $i$ and $j$ index samples within a batch, and $\tau$ is the temperature of the contrastive distribution.

\begin{figure}[ht]
\centering
\includegraphics[width=0.85\columnwidth]{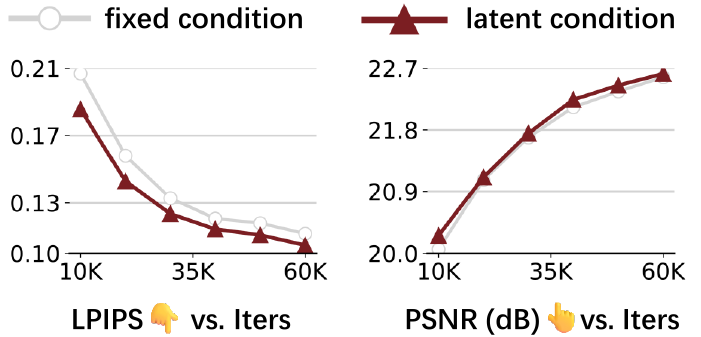}
\caption{Ablation of latent-conditioned guidance.}
\label{fig:cond2}
\end{figure}

During training, latent and text embeddings are co-aligned; at inference, the model relies solely on latent conditioning. Compared to fixed, image-agnostic conditioning~\cite{zhang2025stablecodec}, this approach improves perceptual fidelity and semantic consistency (Fig.~\ref{fig:cond2}). Similar ideas appear in OneDC~\cite{xue2025one}, which uses image tokenizers as supervision. Although the mechanisms differ, both underscore the critical role of condition in diffusion-based reconstruction.

Notably, at extremely low bitrates (e.g., $<0.01$~bpp), the boundary between compression and generation becomes blurred.
As observed in~\cite{keultra}, when visual bits are highly constrained, text tokens may dominate the bit budget. In such scenarios, the latent alone may lack sufficient semantic information, and incorporating auxiliary text priors could further enhance perceptual quality, representing an promising direction for future research.

\begin{figure*}[ht]
\centering
\includegraphics[width=0.95\textwidth]{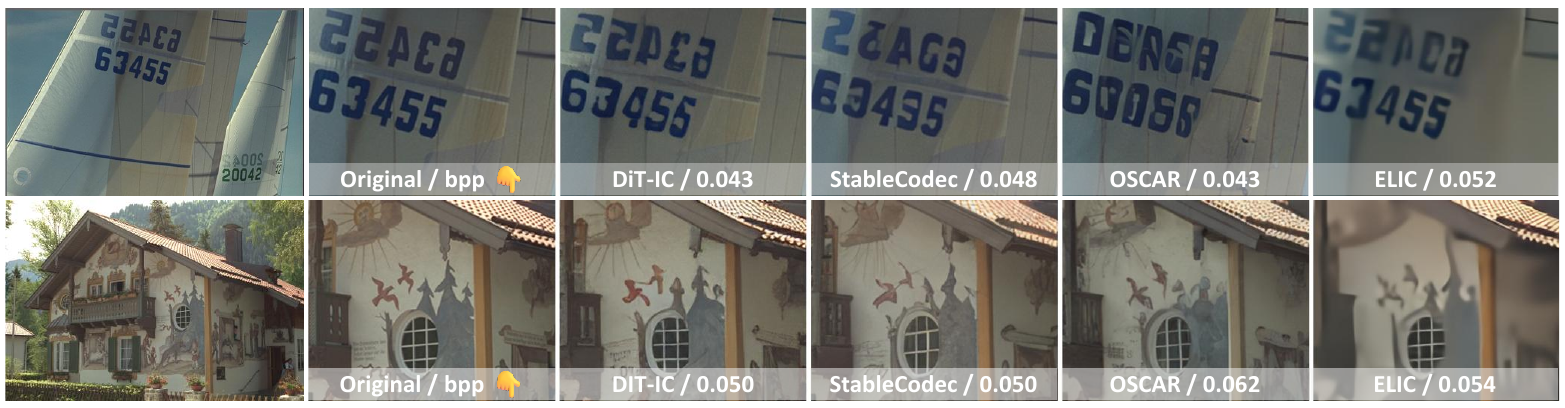}
\caption{\textbf{Visualization comparison.} MSE-optimized ELIC~\cite{he2022elic} suffers from high-frequency detail loss, whereas diffusion-based codecs such as StableCodec~\cite{zhang2025stablecodec} and OSCAR~\cite{guo2025oscar} produce inconsistent semantic content, e.g., incorrect numbers or window panes. In contrast, DiT-IC achieves a more favorable balance between perceptual quality and semantic consistency.}
\label{fig:visual}
\end{figure*}

\subsection{End-to-End Optimization}
To efficiently adapt a pretrained text-to-image diffusion transformer (SANA~\cite{xie2024sana}) for compression, we insert lightweight LoRA~\cite{hu2022lora} adapters, avoiding costly full-model retraining.

We train DiT-IC across a wide range of bitrates using a two-stage \emph{implicit bitrate pruning (IBP)} strategy~\cite{zhang2025stablecodec}: Stage 1 trains the model with a small rate–distortion trade-off $\lambda_\text{base} \in \{0.1, 0.5\}$, relaxing entropy constraints to preserve rich feature representations. Stage 2 fine-tunes the same model with a larger $\lambda_\text{target} \in \{0.5, 1.0, 2.0, 4.0, 8.0, 16.0\}$, progressively tightening bitrate constraints and incorporating adversarial objectives for enhanced perceptual quality.

The overall optimization objective is formulated as:
\begin{align}
&\text{Stage 1:} \min \lambda_\text{base}\mathcal{R} + \mathcal{D} + \mathcal{L}_{align} \\
&\text{Stage 2:} \min \lambda_\text{target}\mathcal{R} + \mathcal{D} + \mathcal{L}_{align} + \lambda_\text{adv} \mathcal{L}_{adv}  \\
&\quad \mathcal{R}(\hat{y},\hat{z}) =  -\log _2 p_{\hat{\mathbf{y}}}(\hat{\mathbf{y}} \mid \hat{\mathbf{z}})-\log _2 p_{\hat{\mathbf{z}}}(\hat{\mathbf{z}}) \\
&\quad \mathcal{D}(x,\hat{x}) = \lambda_\text{1} {MSE} + \lambda_\text{2} {LPIPS} + \lambda_\text{3} {DISTS} \\
&\quad \mathcal{L}_{align}(c,\hat{y}_0) = \lambda_\text{4} \mathcal{L}_{distil} + \lambda_\text{5} \mathcal{L}_{cond}
\label{eq:loss_total}
\end{align}

In practice, we set the LoRA ranks to 32 for the VAE decoder and 64 for the diffusion transformer. For latent–semantic co-alignment, we adopt InternVL~\cite{chen2024internvl, wang2025internvl3_5} as the vision–language backbone, consistent with the original DiT setup. Additional implementation details are provided in the supplementary material.

%% file: sec/4_experiment.tex
\section{Experiment}
\label{sec:experiment}

\subsection{Implementation}
\textbf{Training.}  
We train DiT-IC on a curated dataset of roughly 150K high-quality images with resolutions above $512\times512$, aggregated from CLIC 2020 Professional~\cite{toderici2020clic}, MLIC-Train-100K~\cite{jiang2023mlicpp}, and LSDIR~\cite{li2023lsdir}. 
Training follows a two-stage schedule: the first stage runs for 100K iterations on $256\times256$ patches with a batch size of 32, and the second stage continues for 60K iterations on $512\times512$ patches with a batch size of 16. 
We adopt the AdamW optimizer~\cite{loshchilov2017decoupled} with an initial learning rate of $1\times10^{-4}$, decayed by 50\% at 50\%, 80\%, and 90\% of total iterations.
Consistent with diffusion model practices~\cite{esser2024scaling}, we maintain an exponential moving average (EMA) of model with a decay rate of 0.999.

\textbf{Datasets.}  
We evaluate DiT-IC on three widely used benchmarks: the CLIC 2020 Professional test set~\cite{toderici2020clic}, DIV2K validation set~\cite{agustsson2017ntire}, and Kodak dataset~\cite{Kodak}, following the same protocol as~\cite{zhang2025stablecodec, li2024towards}.
The CLIC 2020 and DIV2K sets contain 428 and 100 images at 2K resolution, while Kodak includes 24 natural images of $768\times512$ pixels.
All evaluations are performed at the original image resolutions without resizing.

\textbf{Metrics.}  
We comprehensively assess the rate–distortion –perception trade-off.
Bitrate is reported in bits per pixel (bpp).
Reconstruction fidelity is measured using PSNR and MS-SSIM~\cite{wang2003multiscale}, and perceptual quality is evaluated with LPIPS~\cite{zhang2018unreasonable} (AlexNet variant by default) and DISTS~\cite{ding2020image}.
Following recent findings~\cite{zhang2025stablecodec, careil2023towards}, we emphasize DISTS as a more reliable indicator of perceptual similarity, particularly in low-bitrate regimes.

\textbf{Baselines.}
To ensure fair and reproducible comparisons, we primarily include open-source methods as baselines.
Some novel approaches, such as \cite{relic2025bridging} and OneDC~\cite{xue2025one}, have not yet released official implementations or publications, and are therefore excluded from our current benchmark for fairness.
We plan to incorporate these methods in future revisions once official resources become available.

All trainings are conducted on two NVIDIA RTX Pro 6000 GPUs, while evaluation and latency benchmarking are performed on an A100 GPU.
Further implementation details are provided in the supplementary material.

\begin{figure*}[htbp]
  \centering
  % 第一行：legend 横跨四列
  \begin{subfigure}{1.0\textwidth}
    \centering
    \includegraphics[width=0.9\textwidth]{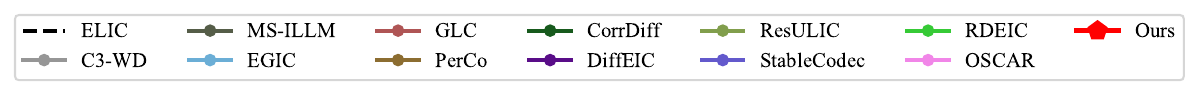}
  \end{subfigure}

  % 第二行（4张图）
  \begin{subfigure}{0.24\textwidth}
    \centering
    \includegraphics[width=\linewidth]{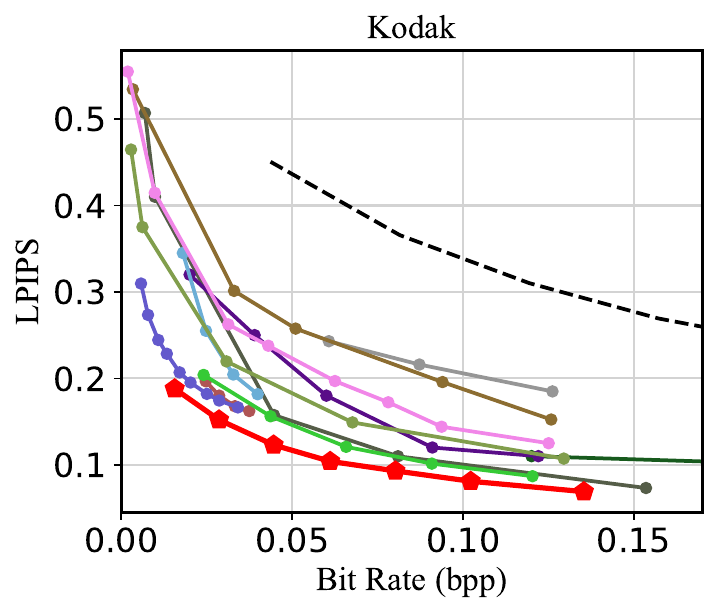}
  \end{subfigure}\hfill
  \begin{subfigure}{0.248\textwidth}
    \centering
    \includegraphics[width=\linewidth]{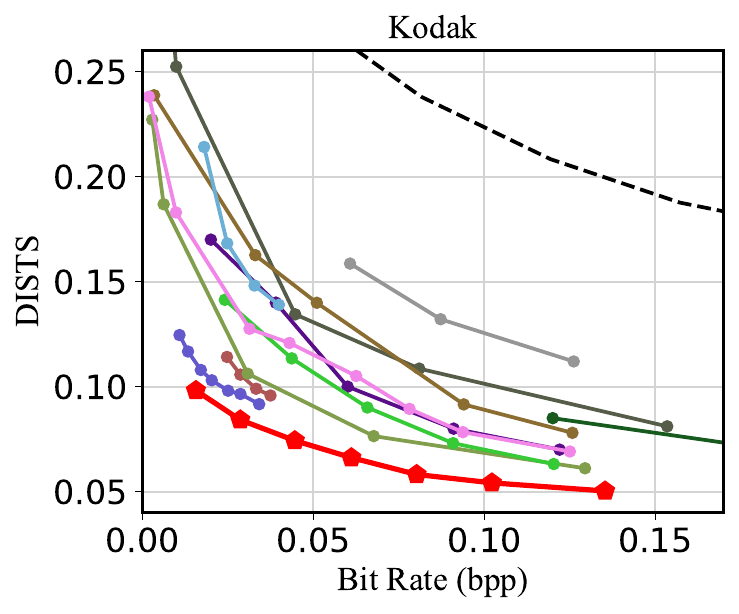}
  \end{subfigure}\hfill
  \begin{subfigure}{0.25\textwidth}
    \centering
    \includegraphics[width=\linewidth]{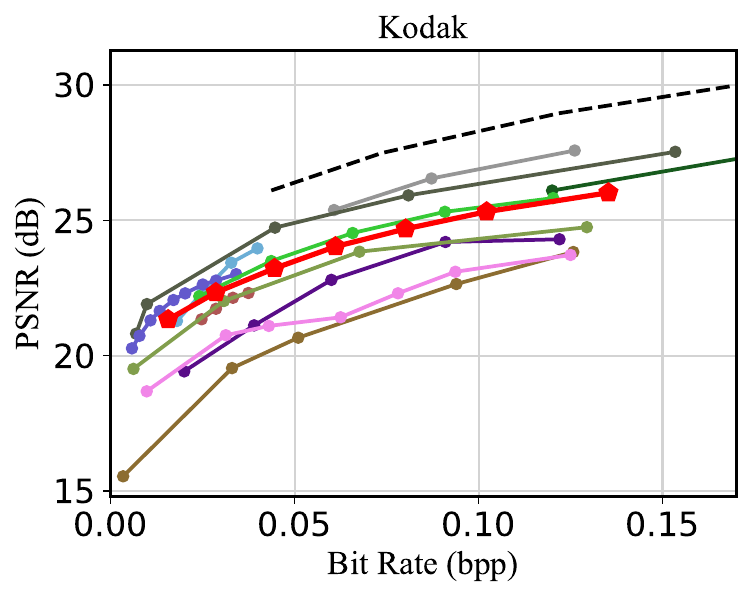}
  \end{subfigure}\hfill
  \begin{subfigure}{0.24\textwidth}
    \centering
    \includegraphics[width=\linewidth]{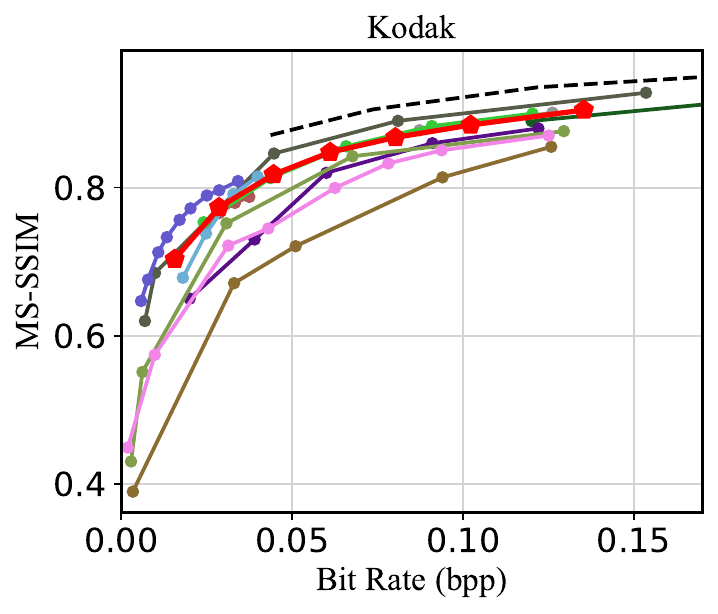}
  \end{subfigure}

  % 第三行（4张图）
  \begin{subfigure}{0.24\textwidth}
    \centering
    \includegraphics[width=\linewidth]{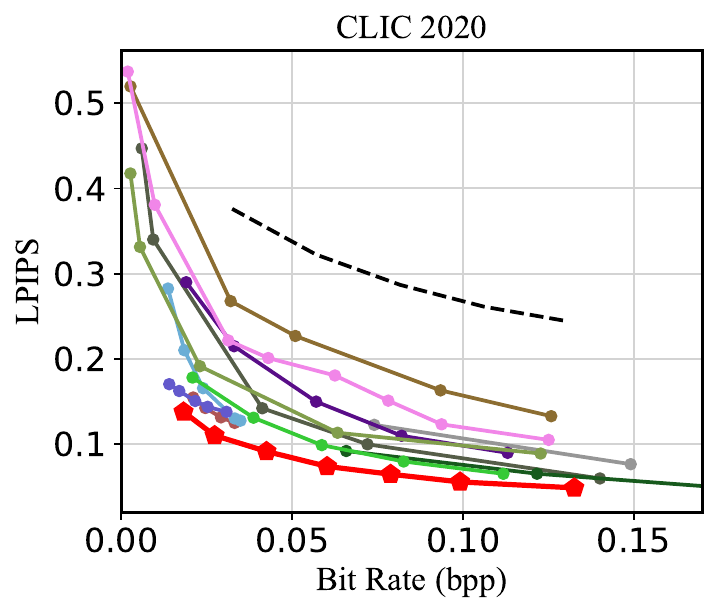}
  \end{subfigure}\hfill
  \begin{subfigure}{0.248\textwidth}
    \centering
    \includegraphics[width=\linewidth]{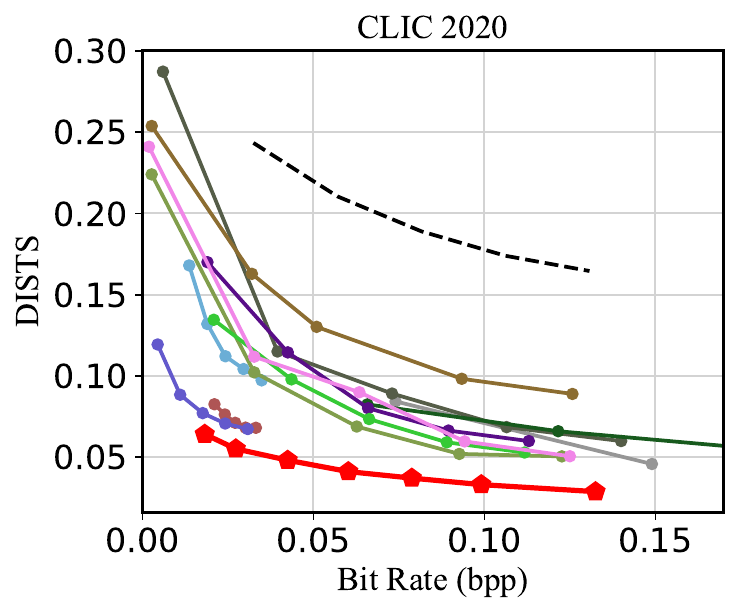}
  \end{subfigure}\hfill
  \begin{subfigure}{0.248\textwidth}
    \centering
    \includegraphics[width=\linewidth]{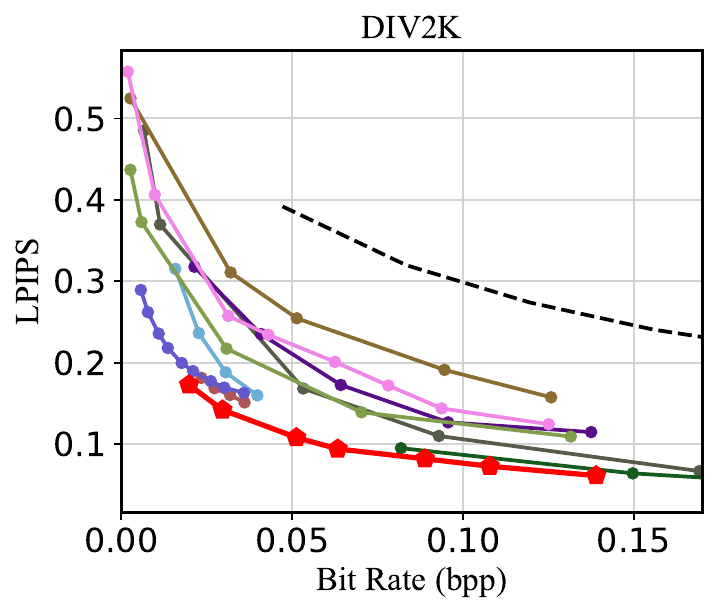}
  \end{subfigure}\hfill
  \begin{subfigure}{0.248\textwidth}
    \centering
    \includegraphics[width=\linewidth]{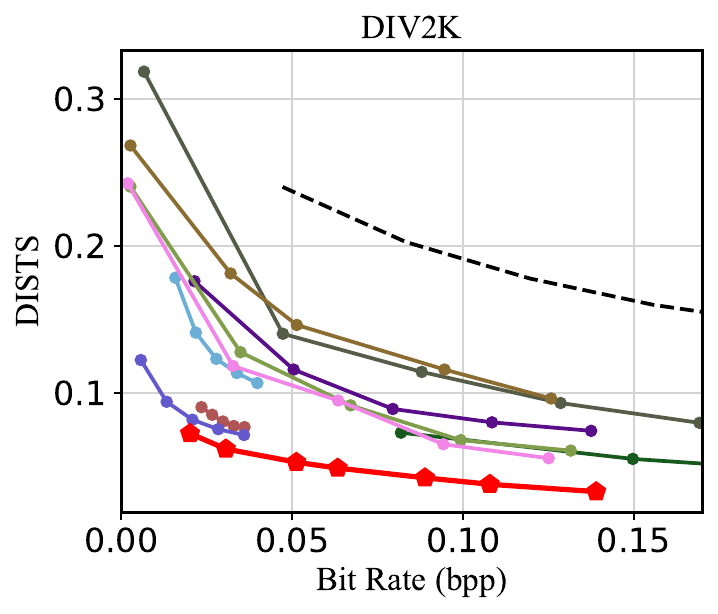}
  \end{subfigure}

  \caption{\textbf{Rate-distortion-perception curve comparisons of different methods on the Kodak, CLIC2020 and DIV2K dataset.}}
  \label{fig:rd_curves}
\end{figure*}

\begin{table*}[htbp]
  \centering
  \renewcommand{\arraystretch}{1.15} % 调整行距
  \setlength{\tabcolsep}{3pt} % 调整列距
  \footnotesize
  \caption{\textbf{Comprehensive comparison with state-of-the-art methods in terms of BD-rate ($\downarrow$)~\cite{bjontegaard2001calculation}.}
“Diff. Reso.” and “Code Reso.” denote the latent resolutions used in the diffusion and coding stages, respectively, where \textit{f} indicates the spatial downsampling factor relative to the pixel domain, and \textit{d} denotes the number of channels.
Latency is measured as the per-image decoding time (for $1024\times1024$ resolution) on a single A100 GPU; $\clubsuit$ marks FP16 inference and $\spadesuit$ marks FP32.
“DiT-IC (baseline)” represents the variant without the proposed alignment strategies.
The best results are highlighted in \textcolor{magenta}{red}, and the second-best in \textcolor{cyan}{blue}.}
  \begin{tabular}{
    l % Methods
    l l c r % diff space, code space, diff steps, Params
    c| % Decoding Latency
    r r r r r r r r r r r r % 12 metric columns (LPIPS/ DISTS × 3 datasets + Avg)
  }
  \toprule
  \multirow{2}[2]{*}{\textbf{Methods}} 
  & \multicolumn{1}{c}{\textbf{Diff.}} & \multicolumn{1}{c}{\textbf{Code}}
  & \multicolumn{1}{c}{\textbf{Diff.}} & \multirow{2}[2]{*}{\textbf{Params}} 
  & \multirow{2}[2]{*}{\textbf{Latency}} 
  & \multicolumn{3}{c}{\textbf{LPIPS $\downarrow$}} & \multicolumn{3}{c}{\textbf{DISTS $\downarrow$}} & \multicolumn{2}{c}{\textbf{Average}} \\
  \cmidrule(lr){7-9} \cmidrule(lr){10-12} \cmidrule(lr){13-14}
  & \multicolumn{1}{c}{\textbf{Reso.}} & \multicolumn{1}{c}{\textbf{Reso.}} 
  & \multicolumn{1}{c}{\textbf{Steps}} &  &  
  & \textbf{Kodak} & \textbf{CLIC} & \textbf{DIV2K} 
  & \textbf{Kodak} & \textbf{CLIC} & \textbf{DIV2K} 
  & \textbf{LPIPS} & \textbf{DISTS} \\
  \midrule
  \textcolor{gray}{\textbf{\textit{INR-based}}} &  &  &  &  &  &  &  &  &  &  &  &  &  \\
  C3-WD (CVPR'25)~\cite{balle2025good} &-  &-  &-  &-  &-  & 12.90 & -47.14 &-  & 62.25 & -46.62 &-  & -17.12 & 7.82 \\
  \midrule
  \textcolor{gray}{\textbf{\textit{VAE-based}}} &  &  &  &  &  &  &  &  &  &  &  &  &  \\
  MS-ILLM (ICML'23)~\cite{muckley2023improving} &-  & f16d256 &-  & 181M & 0.17s & -38.13 & -46.75 & -39.52 & 11.54 & -21.10 & 17.17 & -41.47 & 2.54 \\
  EGIC (ECCV'24)~\cite{korber2024egic} &-  & f16d320 &-  & 37M &  & -47.93 & -67.75 & -61.75 & 0.07 & -60.69 & -61.19 & -59.14 & -40.60 \\
  GLC (TCSVT'25)~\cite{qi2025generative} &-  & f16d256 &-  & 105M & 0.18s & -72.04 & -78.57 & -75.90 & -63.21 & -84.33 & -87.60 & -75.50 & -78.38 \\
  \midrule
  \textcolor{gray}{\textbf{\textit{Diffusion-based}}} &  &  &  &  &  &  &  &  &  &  &  &  &  \\
  PerCo (ICLR'24)~\cite{careil2023towards, korber2024perco} & f8d4 & f8-64d320 & 20 & 4.3B & 8.8s & 0.00 & 0.00 & 0.00 & 0.00 & 0.00 & 0.00 & 0.00 & 0.00 \\
  CorrDiff (ICML'24)~\cite{ma2024correcting} & f0d3 & f16d320 & 8 & 73M &  & -69.76 & -72.94 & -73.08 & 13.90 & -57.71 & -67.32 & -71.93 & -37.04 \\
  DiffEIC (TCSVT'24)~\cite{li2024towards} & f8d4 & f16d320 & 50 & 1.0B & 12.4s & -33.91 & -40.47 & -34.03 & -25.37 & -36.04 & -39.76 & -36.14 & -33.72 \\
  ResULIC (ICML'25)~\cite{keultra} &f8d4 & f32d192 & 4 & 12.3B & 0.83s & -57.39 & -66.50 & -62.93 & -65.31 & -68.64 & -62.96 & -62.27 & -65.64 \\
  StableCodec (ICCV'25)~\cite{zhang2025stablecodec} & f8d4 & f64d320 & 1 & 1.5B & 0.34s & \textcolor{cyan}{\textbf{-78.34}} & \textcolor{cyan}{\textbf{-80.21}} & \textcolor{cyan}{\textbf{-79.02}} & \textcolor{cyan}{\textbf{-70.48}} & \textcolor{cyan}{\textbf{-90.24}} & \textcolor{cyan}{\textbf{-91.14}} & \textcolor{cyan}{\textbf{-79.19}} & \textcolor{cyan}{\textbf{-83.95}} \\
  RDEIC (TCSVT'25)~\cite{li2025rdeic} & f8d4 & f16d256 & 5 & 1.0B & 1.5s & -67.99 & -71.95 &-  & -39.02 & -53.05 &-  & -69.97 & -46.04 \\
  OSCAR (NeurIPS'25)~\cite{guo2025oscar} & f8d4 & f8-64d320 & 1 & 987M & \textcolor{cyan}{\textbf{0.32s}} & -17.80 & -14.74 & -24.58 & -46.18 & -59.82 & -69.15 & -19.04 & -58.38 \\
  % OneDC (NeurIPS'25)~\cite{xue2025one} & f8d4 & f16d128 & 1 & 1.3B & 0.34s & \textcolor{magenta}{\textbf{-82.35}} & \textcolor{magenta}{\textbf{-87.31}} & \textcolor{magenta}{\textbf{-84.40}} & \textcolor{cyan}{\textbf{-70.99}} & \textcolor{cyan}{\textbf{-92.62}} & \textcolor{cyan}{\textbf{-91.27}} & \textcolor{magenta}{\textbf{-84.69}} & \textcolor{cyan}{\textbf{-84.96}} \\
  \midrule
  \rowcolor{cyan!10}\graycell{DiT-IC (baseline)} & \graycell{f32d32} & \graycell{f64d320} & \graycell{1} & \graycell{990M} & \textcolor{magenta}{\textbf{0.27s} $\spadesuit$} & \graycell{-62.36} & \graycell{-68.15} & \graycell{-64.50} & \graycell{-54.20} & \graycell{-73.58} & \graycell{-73.78} & \graycell{-65.00} & \graycell{-67.19} \\
  \rowcolor{cyan!10}\textbf{DiT-IC (Ours)} &f32d32  &f64d320  & 1 & 1.0B & \textcolor{magenta}{\textbf{(0.15s) $\clubsuit$}} & \textcolor{magenta}{\textbf{-81.11}} & \textcolor{magenta}{\textbf{-86.73}} & \textcolor{magenta}{\textbf{-83.11}} & \textcolor{magenta}{\textbf{-75.95}} & \textcolor{magenta}{\textbf{-94.45}} & \textcolor{magenta}{\textbf{-93.25}} & \textcolor{magenta}{\textbf{-83.65}} & \textcolor{magenta}{\textbf{-87.88}} \\
  \bottomrule
  \end{tabular}
  \label{tab:rd}
\end{table*}

\subsection{Main Results}

\textbf{Quantitative Performance Comparison.}
Fig.~\ref{fig:rd_curves} and Table~\ref{tab:rd} provide a comprehensive comparison of rate–distortion efficiency and perceptual fidelity across representative image compression methods.
Conventional VAE-based codecs~\cite{muckley2023improving, korber2024egic, qi2025generative} offer fast inference but are constrained by limited representational capacity, resulting in suboptimal visual realism.
In contrast, early diffusion-based codecs~\cite{korber2024perco, ma2024correcting, li2024towards, keultra, li2025rdeic} markedly improve perceptual quality, yet suffer from excessive computational overhead—typically requiring 4–50 iterative denoising steps and over one second of decoding time per image.
Recent one-step diffusion codecs~\cite{zhang2025stablecodec, guo2025oscar, xue2025one} alleviate this issue by accelerating reconstruction, but they largely depend on pretrained Stable Diffusion backbones.
Due to their U-Net-based architectures, these models struggle to operate in deeper latent domains, leading to persistent inefficiency.
In contrast, DiT-IC performs diffusion entirely within a deeper latent space using a transformer-based architecture, effectively reducing complexity while preserving expressive capacity.
These results highlight the advantages of our diffusion transformer paradigm in achieving efficient and high-fidelity image compression.

\textbf{Qualitative Visualization.}
While MSE-optimized codecs such as ELIC~\cite{he2022elic} maintain overall semantic consistency, they tend to produce overly smooth textures that deviate from human perceptual preference.
Diffusion-based codecs, on the other hand, enhance perceptual realism but often introduce color shifts or semantic distortions—undesirable in compression scenarios where fidelity is critical.
As illustrated in Fig.~\ref{fig:visual}, StableCodec~\cite{zhang2025stablecodec} misrepresents fine-grained details such as the boat number and window panes, reflecting unstable semantic reconstruction.
In contrast, DiT-IC achieves a better trade-off between perceptual quality and structural fidelity under comparable entropy constraints.
Additional visualizations are provided in the supplementary material.

\begin{table}[tbp]
  \centering
  \footnotesize
  \setlength{\tabcolsep}{5pt} % 调整列距
  \caption{Ablation study results measured by BD-rate $\downarrow$~\cite{bjontegaard2001calculation}.}
    \begin{tabular}{lrrrr}
    \toprule
          & PSNR  & MS-SSIM & LPIPS & DISTS \\
    \midrule
    DiT-IC & 0.00\% & 0.00\% & 0.00\% & 0.00\% \\
    \midrule
    \multicolumn{5}{c}{\textcolor{gray}{\textit{\textbf{Loss function}}}} \\
    \midrule
    -w/o $\mathcal{L}_{adv}$ & -37.10\% & -21.54\% & -2.27\% & -1.80\% \\
    -w/o DISTS & -2.15\% & -1.30\% & -1.83\% & 5.69\% \\
    \midrule
    \multicolumn{5}{c}{\textcolor{gray}{\textit{\textbf{Training Strategies}}}} \\
    \midrule
    DiT from scratch & 16.80\% & 13.41\% & 22.00\% & 32.45\% \\
    \ -VAE/DiT rank 16/16 & 8.24\% & 6.43\% & 12.77\% & 13.92\% \\
    \ -VAE/DiT rank 32/32 & 3.10\% & 2.76\% & 5.31\% & 5.56\% \\
    \ -full finetuning & 3.52\% & 3.21\% & 7.95\% & 8.05\% \\
    \bottomrule
    \end{tabular}%
  \label{tab:ablation}%
\end{table}%

\subsection{Ablations and Discussion}
Ablation results, are trained on $256\times256$ images with 60K iterations, including Table~\ref{tab:ablation} and those reported in Sec.~\ref{sec:method}.

\textbf{Loss Formulations.}  
Introducing the adversarial term $\mathcal{L}_{adv}$ yields perceptually sharper and more realistic reconstructions, aligning better with human visual preference.
However, this comes at the cost of slight degradation in quantitative metrics. Given its substantial improvement in perceptual realism, we retain $\mathcal{L}_{adv}$ in the final objective.
Additionally, incorporating the DISTS term further strengthens correlation with human perception, especially under low-bitrate regimes, albeit with a minor trade-off in distortion-oriented metrics.

\textbf{Training Strategies.}  
We initialize DiT-IC using pretrained SANA weights~\cite{xie2024sana}.
Training the entire model from scratch leads to noticeable performance degradation, likely due to limited training data scale.
We further examine different LoRA configurations for both the VAE and DiT modules.
A rank setting of 32/64 achieves the best balance between adaptation capacity and stability, whereas full fine-tuning slightly degrades performance—possibly because large-scale parameter updates distort the pretrained distribution manifold when trained under small-batch conditions.

% Table generated by Excel2LaTeX from sheet 'Train'
\begin{table}[tbp]
  \centering
  \footnotesize
  \setlength{\tabcolsep}{3pt} % 调整列距
  \caption{Runtime latency (s) comparison in FP32 precision.}
    \begin{tabular}{lllll}
    \toprule
          &Reso.       & VAE   & Codec & Diffusion \\
    \midrule
    StableCodec &-  & SD-VAE~\cite{sauer2024adversarial} & f8 $\rightarrow$ f64 & U-Net \\
    DiT-IC  &-  & SANA-VAE~\cite{xie2024sana} & f32 $\rightarrow$ f64 & DiT \\
    \midrule
    StableCodec & $1024^2$ & \multicolumn{1}{l}{0.19} & \multicolumn{1}{l}{0.04} & \multicolumn{1}{l}{0.11} \\
    DiT-IC  & $1024^2$ & 0.21 \redcell{(+11\%)} & 0.008  \bluecell{(-30\%)} & 0.055  \bluecell{(-50\%)} \\
    \midrule
    StableCodec & $2048^2$ & \multicolumn{1}{l}{0.82} & \multicolumn{1}{l}{0.05} & \multicolumn{1}{l}{0.8} \\
    DiT-IC  & $2048^2$ & 0.85 \redcell{(+4\%)} & 0.012 \bluecell{(-76\%)} & 0.12 \bluecell{(-85\%)} \\
    \midrule
    StableCodec & $4096^2$ & \multicolumn{1}{l}{85} & \multicolumn{1}{l}{0.13} & \multicolumn{1}{l}{10.3} \\
    DiT-IC & $4096^2$ & 3.3  \bluecell{(-96\%)} & 0.022  \bluecell{(-83\%)} & 0.47 \bluecell{(-95\%)} \\
    \bottomrule
    \end{tabular}%
  \label{tab:latency}%
\end{table}%

\textbf{Complexity.}  
Table~\ref{tab:latency} compares the latency (in second) of DiT-IC and StableCodec~\cite{zhang2025stablecodec}.
For fair comparison, tiled VAE coding is disabled in both models.
Despite adopting single-step diffusion in both frameworks, DiT-IC achieves consistently lower runtime, particularly at higher resolutions.
At $1024^2$ and $2048^2$, diffusion latency is reduced by 50\%–85\%, and overall time by up to 76\%, owing to its operation in a deeper latent space.
Notably, StableCodec exhibits a sharp latency surge at $4096^2$, likely caused by fragmented GPU computation and excessive memory transfer overhead.
In contrast, DiT-IC maintains stable scalability and efficient high-resolution reconstruction, demonstrating strong hardware adaptability.

%% file: sec/5_conclusion.tex
\section{Conclusion}
\label{sec:conclusion}
We presented \textbf{DiT-IC}, an Aligned Diffusion Transformer for efficient image compression.
By shifting the diffusion process into deeply compressed latent domains, DiT-IC effectively mitigates the inherent inefficiency of diffusion sampling.
Through variance-guided reconstruction flow, self-distillation alignment, and latent-conditioned guidance, it aligns pretrained diffusion transformers toward the compression objective, enabling one-step, high-fidelity reconstruction within a $32\times$ latent space.
Extensive experiments demonstrate that DiT-IC achieves state-of-the-art rate–distortion trade-offs, up to 30× faster decoding, and significantly reduced memory cost compared with existing diffusion-based codecs.
We believe this alignment perspective will inspire future research on generative compression and efficient visual representation learning.

%% file: sec/X_suppl.tex
\clearpage
\setcounter{page}{1}
\maketitlesupplementary

\begin{figure*}[ht]
    \centering
    \includegraphics[width=0.95\linewidth]{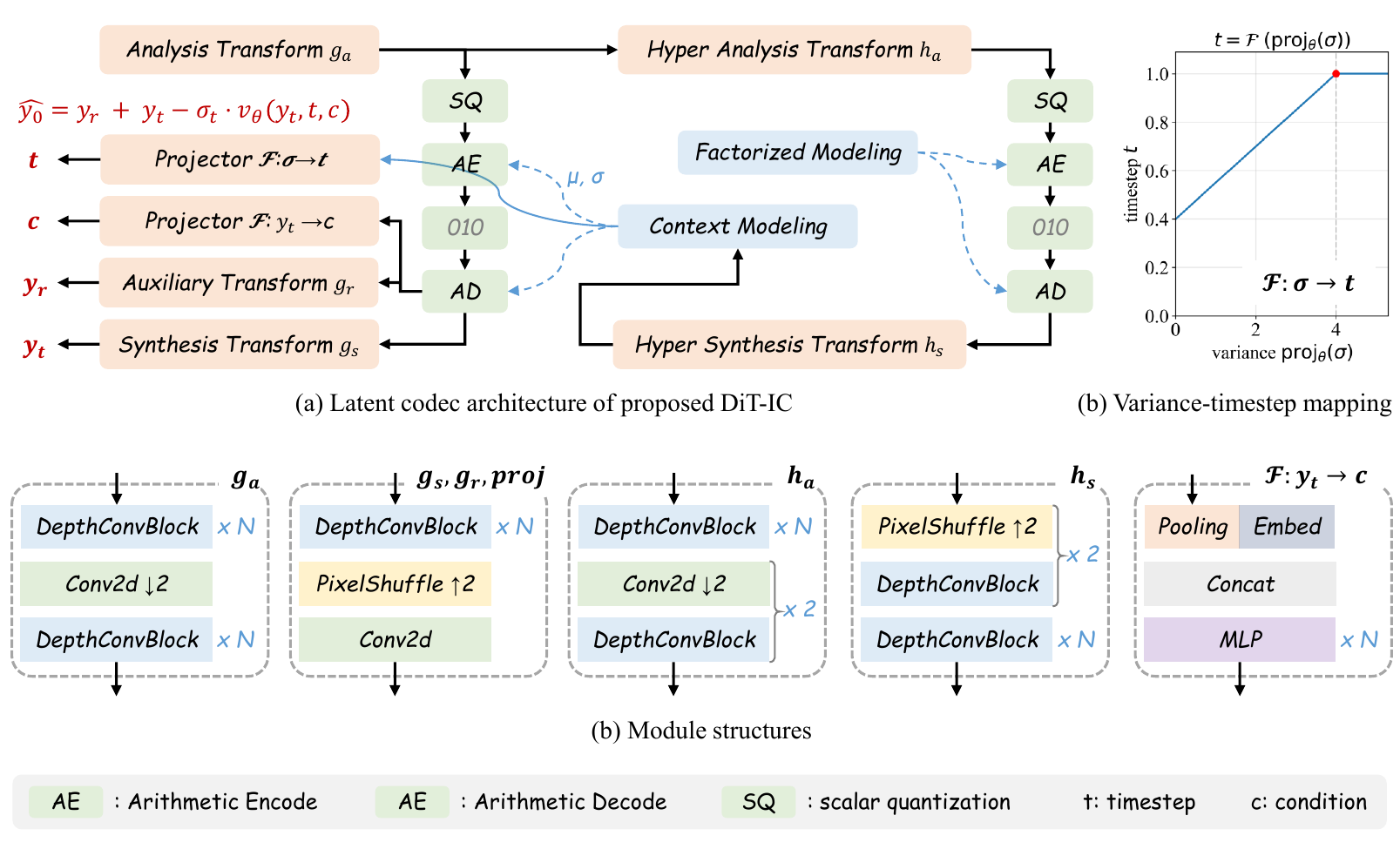}
    \caption{Overall architecture of our model. The entropy model is based on the hyperprior framework and an autoregressive context model similar to StableCodec~\cite{zhang2025stablecodec}, but replaces heavy components with lightweight \textit{DepthConvBlocks}~\cite{jia2025towards}.}
    \label{fig:arch2}
\end{figure*}

\begin{figure*}[ht]
    \centering
    \includegraphics[width=0.95\linewidth]{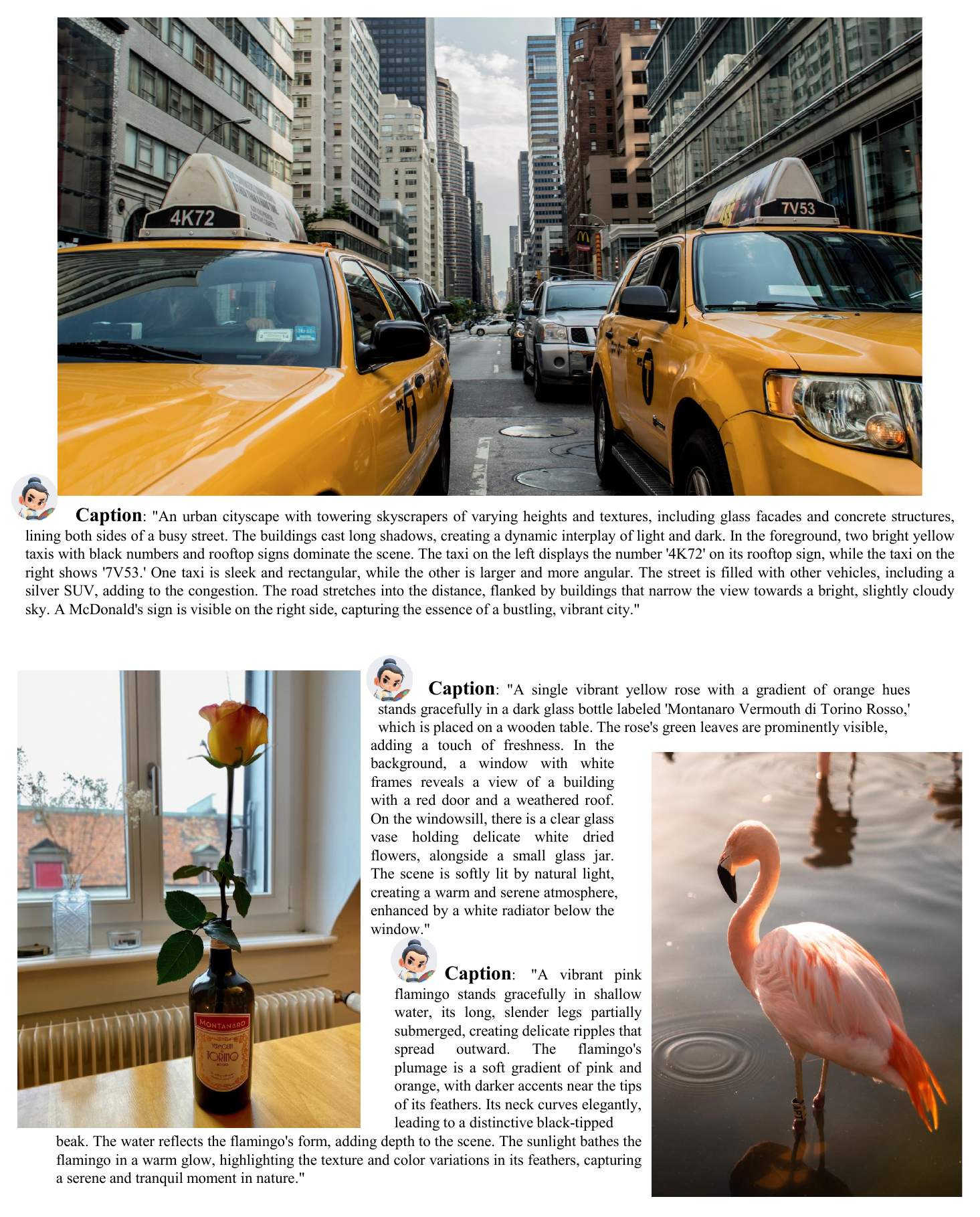}
    \caption{Illustrative VLM-generated captions used for semantic conditioning.}
    \label{fig:caption}
\end{figure*}

\section{Method Details}

\textbf{Model Architecture.}
The overall architecture is illustrated in Fig.~\ref{fig:arch2}. Our entropy model follows the classical hyperprior framework and further incorporates the autoregressive context model introduced in StableCodec~\cite{zhang2025stablecodec}. Different from StableCodec, we replace the original context modules with a lightweight \textit{DepthConvBlock}~\cite{jia2025towards}, which significantly reduces computational complexity while preserving effective spatial–channel context modeling capability. Given the quantized latent representation $\hat{\mathbf{z}}$, the autoregressive module predicts the Gaussian distribution parameters $(\boldsymbol{\mu}, \boldsymbol{\sigma})$ via a 4-step autoregressive procedure. These parameters are then fed into an arithmetic coder to convert quantized symbols into a bitstream during encoding, or to reconstruct symbols from the bitstream during decoding.

\textbf{Resolution Generalization.}
DiT-IC adopts a Diffusion Transformer without positional encoding (NoPE)~\cite{kazemnejad2023impact}, avoiding positional extrapolation issues commonly encountered in standard Transformers. By removing positional embeddings, the model does not bind representations to fixed spatial indices, thereby improving length and resolution generalization. Although trained on small patches, the model generalizes reliably to higher resolutions at inference time without architectural modification.

\textbf{Self-Distillation Alignment.}
The key idea is to collapse multi-step diffusion supervision into a self-aligned single-step objective without introducing an external teacher model. We adopt alignment-style objectives to approximate diffusion behavior under a one-step formulation, and term this strategy \textit{Self-Distillation Alignment} to distinguish it from conventional teacher–student distillation methods. This formulation preserves diffusion-style supervision while avoiding additional model overhead.

\textbf{Variance–Timestep Mapping.}
As shown in Fig. 4, the predicted variance exhibits a strong correlation with compressed noise (with cosine similarity up to 0.94). From a variational inference perspective, higher latent variance corresponds to higher conditional entropy and greater reconstruction uncertainty, which manifests as stronger noise components. This observation motivates a monotonic variance-to-timestep mapping strategy: larger variance is mapped to a larger diffusion timestep, implying stronger denoising. Consequently, entropy modeling and timestep prediction are naturally aligned. Empirically, blocking gradients from the $\mathcal{F}:\sigma \rightarrow t$ branch results in negligible bitrate change, indicating that joint optimization introduces minimal conflict between compression and diffusion objectives.

\textbf{Distortion–Perception Trade-off.} 
Under a fixed information rate, distortion and perceptual quality cannot be simultaneously optimized, as established by the rate–distortion–perception trade-off principle~\cite{blau2019rethinking, blau2018perception, niu2023conditional}. DiT-IC adheres to this information-theoretic constraint, which explains why perceptual optimization may lead to reduced PSNR. The trade-off is controlled by the weighting parameter $\lambda$ in Eq.~(10). In practice, sweeping $\lambda$ produces smooth distortion–perception operating curves, allowing flexible control over reconstruction fidelity and perceptual realism.

\begin{table*}[htbp]
  \centering
  \footnotesize
  \caption{Quantitative perceptual comparison between DiT-IC and StableCodec ($\lambda=2.0$) at similar bitrates ($\sim$0.03–0.04 bpp). We report a suite of perceptual metrics including FID, KID, NIQE, CLIPIQA and MUSIQ. Lower is better for FID/KID/NIQE, and higher is better for the remaining metrics. DiT-IC consistently outperforms StableCodec across most datasets and metrics, with a slight exception on KID for CLIC2020.  The best results are highlighted in \textcolor{magenta}{\textbf{red}}.}
    \begin{tabular}{lrrrrrrrrrr}
    \toprule
    Datasets & \multicolumn{2}{c}{Kodak} & \multicolumn{2}{c}{CLIC 2020} & \multicolumn{2}{c}{DIV2K} & \multicolumn{2}{c}{Average}     & \multicolumn{2}{c}{Avg. Difference}  \\
    \midrule
    Methods & \multicolumn{1}{l}{DiT-IC} & \multicolumn{1}{l}{StableCodec} & \multicolumn{1}{l}{DiT-IC} & \multicolumn{1}{l}{StableCodec} & \multicolumn{1}{l}{DiT-IC} & \multicolumn{1}{l}{StableCodec} & \multicolumn{1}{l}{DiT-IC} & \multicolumn{1}{l}{StableCodec} & \multicolumn{1}{l}{$|\Delta| \uparrow$} & \multicolumn{1}{l}{$|\Delta| (\%) \uparrow$} \\
    \midrule
    FID $\downarrow$   &     -  &   -    & \textcolor{magenta}{\textbf{3.750}}  & 3.940  & \textcolor{magenta}{\textbf{8.650}}  & 10.350  & \textcolor{magenta}{\textbf{6.200}}  & 7.145  & 0.945 & 13.23\% \\
    KID $\downarrow$  &   -    &  -     & 0.00083  & \textcolor{magenta}{\textbf{0.00066}}  & \textcolor{magenta}{\textbf{0.00060}}  & 0.00080  & \textcolor{magenta}{\textbf{0.00072}}  & 0.00073  & 0.00002 & 2.06\% \\
    NIQE $\downarrow$ & \textcolor{magenta}{\textbf{3.099}}  & 3.557  & \textcolor{magenta}{\textbf{3.833}}  & 4.459  & \textcolor{magenta}{\textbf{3.270}}  & 3.603  & \textcolor{magenta}{\textbf{3.400}}  & 3.873  & 0.473 & 12.21\% \\
    CLIPIQA $\uparrow$ & \textcolor{magenta}{\textbf{0.735}}  & 0.716  & \textcolor{magenta}{\textbf{0.582}}  & 0.531  & \textcolor{magenta}{\textbf{0.626}}  & 0.570  & \textcolor{magenta}{\textbf{0.648}}  & 0.606  & 0.042  & 6.92\% \\
    MUSIQ $\uparrow$ & \textcolor{magenta}{\textbf{74.494}}  & 73.177  & \textcolor{magenta}{\textbf{60.606}}  & 58.663  & \textcolor{magenta}{\textbf{65.818}}  & 63.822  & \textcolor{magenta}{\textbf{66.972}}  & 65.221  & 1.752  & 2.69\% \\
    \bottomrule
    \end{tabular}%
  \label{tab:result2}%
\end{table*}%

\begin{figure*}[htbp]
  \centering
  % 第一行：legend 横跨四列
  \begin{subfigure}{1.0\textwidth}
    \centering
    \includegraphics[width=0.9\textwidth]{figs/legend_only.pdf}
  \end{subfigure}

  % 第二行（4张图）
  \begin{subfigure}{0.24\textwidth}
    \centering
    \includegraphics[width=\linewidth]{figs/Kodak-LPIPS.pdf}
  \end{subfigure}\hfill
  \begin{subfigure}{0.248\textwidth}
    \centering
    \includegraphics[width=\linewidth]{figs/Kodak-DISTS.pdf}
  \end{subfigure}\hfill
  \begin{subfigure}{0.248\textwidth}
    \centering
    \includegraphics[width=\linewidth]{figs/Kodak-PSNR.pdf}
  \end{subfigure}\hfill
  \begin{subfigure}{0.24\textwidth}
    \centering
    \includegraphics[width=\linewidth]{figs/Kodak-SSIM.pdf}
  \end{subfigure}

  % 第三行（4张图）
  \begin{subfigure}{0.24\textwidth}
    \centering
    \includegraphics[width=\linewidth]{figs/CLIC-LPIPS.pdf}
  \end{subfigure}\hfill
  \begin{subfigure}{0.248\textwidth}
    \centering
    \includegraphics[width=\linewidth]{figs/CLIC-DISTS.pdf}
  \end{subfigure}\hfill
  \begin{subfigure}{0.248\textwidth}
    \centering
    \includegraphics[width=\linewidth]{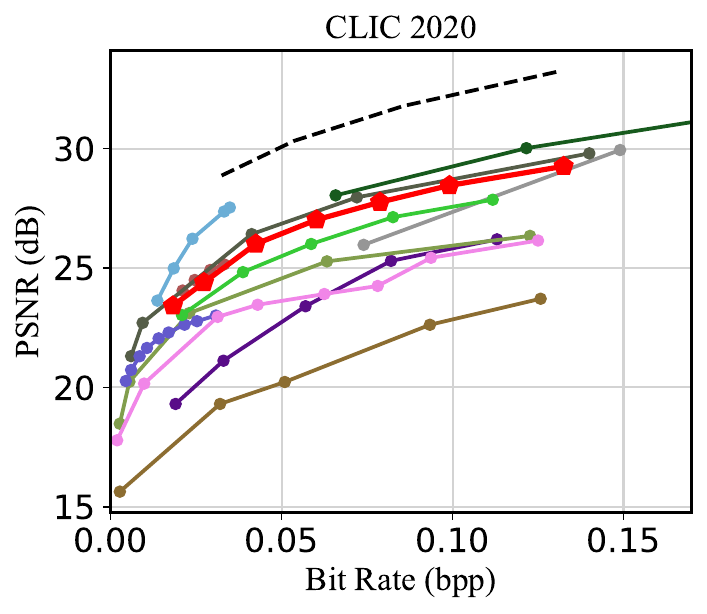}
  \end{subfigure}\hfill
  \begin{subfigure}{0.24\textwidth}
    \centering
    \includegraphics[width=\linewidth]{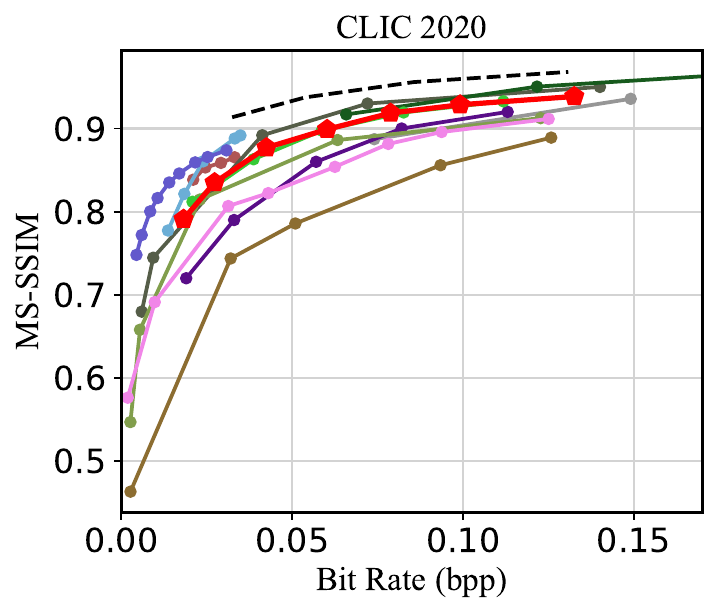}
  \end{subfigure}

  % 第四行（4张图）
  \begin{subfigure}{0.24\textwidth}
    \centering
    \includegraphics[width=\linewidth]{figs/DIV-LPIPS.pdf}
  \end{subfigure}\hfill
  \begin{subfigure}{0.248\textwidth}
    \centering
    \includegraphics[width=\linewidth]{figs/DIV-DISTS.pdf}
  \end{subfigure}\hfill
  \begin{subfigure}{0.248\textwidth}
    \centering
    \includegraphics[width=\linewidth]{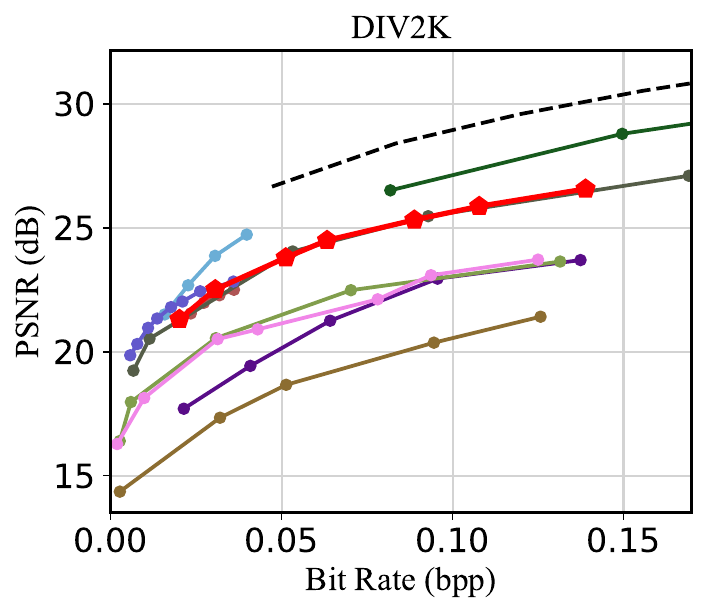}
  \end{subfigure}\hfill
  \begin{subfigure}{0.24\textwidth}
    \centering
    \includegraphics[width=\linewidth]{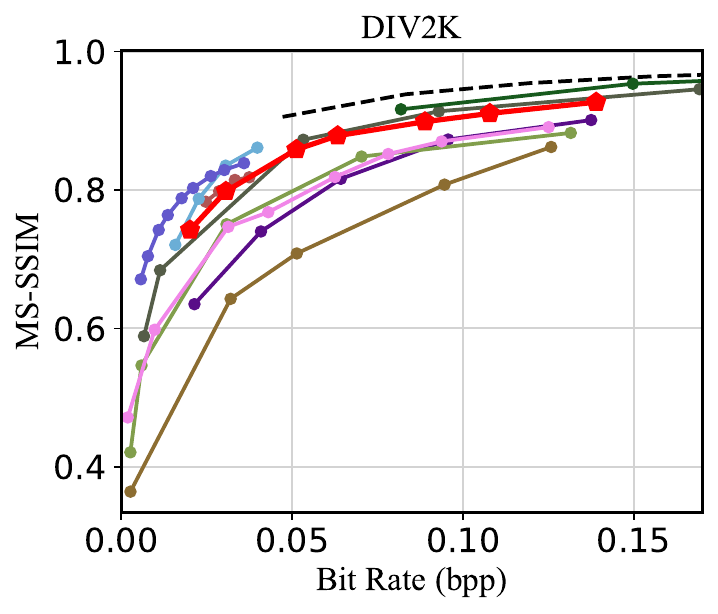}
  \end{subfigure}

  \caption{Detailed Rate-distortion-perception curve comparisons of different methods on the Kodak, CLIC2020 and DIV2K dataset.}
  \label{fig:rd_curves_all}
\end{figure*}

\section{Captions Generated by a VLM}
To avoid manual annotation and ensure scalable supervision, we employ a Vision–Language Model (VLM) to automatically generate semantic captions for training. Specifically, we adopt InternVL~\cite{chen2024internvl, wang2025internvl3_5}, which is consistent with the captioning pipeline used in the original text-to-image DiT-SANA~\cite{xie2024sana} pretraining. Representative caption examples produced by the VLM are shown in Fig.~\ref{fig:caption}.

\section{More Implementation Details}
Our DiT-IC model is trained on two NVIDIA RTX Pro 6000 GPUs using PyTorch~2.8.0 and CUDA~12.8. For fair comparison, we reproduce several open-source baselines within the same environment to obtain detailed results. Due to differences in software versions and numerical kernels, minor deviations from the originally reported numbers may occur.

The training consists of two stages. In Stage~2, we initially disable the adversarial loss by setting $\lambda_{\text{adv}} = 0$, and enable it only after 30\% of iterations to stabilize optimization. We also gradually anneal the contrastive co-alignment loss that aligns latent embeddings with text embeddings, controlled by a temperature parameter~$\tau$. This loss is used only during the initial 30\% of Stage~2 to provide early semantic guidance while avoiding unstable or noisy text-driven updates in later iterations.

After the two-stage training, the model typically reaches a stable convergence point. At this stage, the Self-Distillation Alignment module becomes less essential, and jointly finetuning the entire model—including the encoder—could potentially yield further improvements. Although encoder finetuning is not included in this work, exploring this unified training strategy remains a promising direction for future research.

\section{Complexity}
\textbf{Training Complexity.}
Multi-stage training is commonly adopted in diffusion-based codecs (e.g., StableCodec, OneDC, and ResULIC), often involving external teacher inference or multi-step diffusion supervision. In contrast, our training pipeline is strictly sequential and does not require additional teacher models or iterative diffusion sampling during optimization. In practice, the model converges within approximately 3 days on two NVIDIA A100 GPUs.

\textbf{Memory Usage.}
The reported 16GB memory footprint corresponds to full-frame 2K decoding without tiling. When using $1024 \times 1024$ tiled decoding, peak memory consumption decreases to below \textbf{7GB} without any observable quality degradation. Employing smaller tiles can further reduce memory usage if necessary. Moreover, applying INT8 quantization lowers memory consumption to approximately \textbf{4GB}, making deployment feasible on consumer-grade GPUs.

\section{Quantitative Evaluation}
\textbf{Rate-Distortion Curves.}
In Fig.~\ref{fig:rd_curves_all}, we present full rate–distortion curves on Kodak~\cite{Kodak}, CLIC 2020~\cite{toderici2020clic}, and DIV2K~\cite{agustsson2017ntire} as a supplement to Fig.~11 of main paper. As discussed in main paper, pixel-level metrics such as PSNR and MS-SSIM exhibit notable limitations~\cite{ding2020image, careil2023towards, zhang2025stablecodec}. These metrics primarily emphasize pixel fidelity rather than semantic consistency or perceptual realism, making them less suitable for evaluating compression performance in the ultra–low bitrate regime.

\begin{figure*}[tbp]
  \centering
  \begin{subfigure}{0.24\textwidth}
    \centering
    \includegraphics[width=\linewidth]{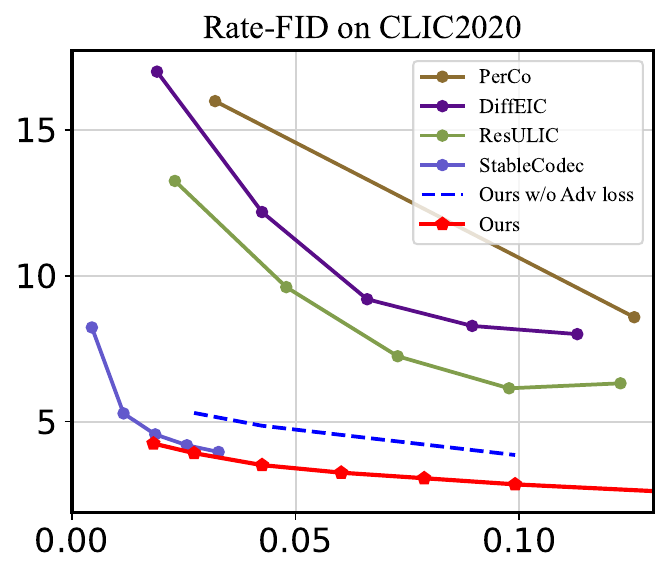}
  \end{subfigure}
  \begin{subfigure}{0.24\textwidth}
    \centering
    \includegraphics[width=\linewidth]{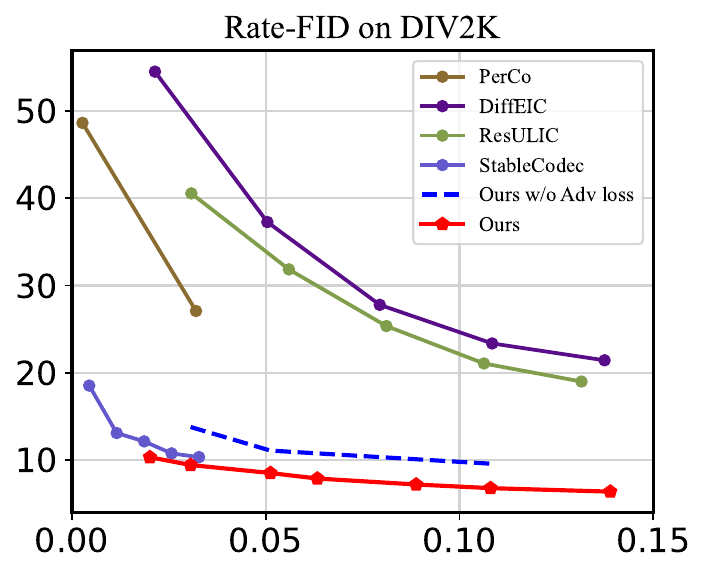}
  \end{subfigure}
  \begin{subfigure}{0.24\textwidth}
    \centering
    \includegraphics[width=\linewidth]{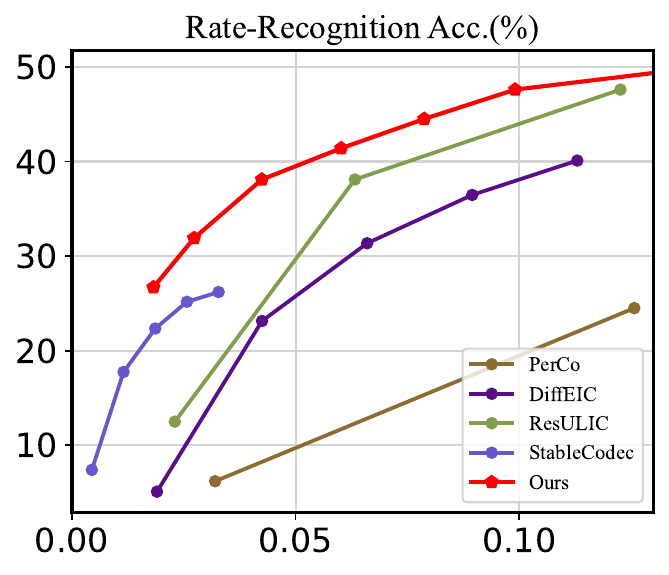}
  \end{subfigure}
  \caption{DiT-IC achieves superior FID and Semantic accuracy.}
  \label{fig:fid}
\end{figure*}

\textbf{Semantic Study.}
We further evaluate semantic fidelity using the OCRBench v2 evaluation pipeline~\cite{fu2024ocrbench}. This protocol measures high-level semantic consistency by applying a unified OCR-based recognition framework to reconstructed images and comparing semantic accuracy against ground truth. Unlike pixel-level metrics, this evaluation directly assesses whether compressed reconstructions preserve semantically meaningful content. As shown in Fig.~\ref{fig:fid} (right), DiT-IC maintains strong semantic consistency, indicating that the perceptual enhancement does not compromise high-level semantic integrity.

\textbf{User Study.}
We conduct a large-scale user study with 61 participants to evaluate perceptual realism. Each participant is presented with randomized pairwise comparisons among ResULIC, PerCo, StableCodec, OSCAR, and DiT-IC at matched bitrates, and is asked to select the visually more realistic reconstruction. The aggregated preference scores are $8.2\%, 1.0\%, 27.5\%, 6.5\%, \mathbf{56.8\%}$ for ResULIC, PerCo, StableCodec, OSCAR, and DiT-IC, respectively. DiT-IC receives the highest preference by a substantial margin, demonstrating its clear advantage in perceptual realism under controlled bitrate settings.

\begin{figure*}[ht]
    \centering
    \includegraphics[width=0.97\linewidth]{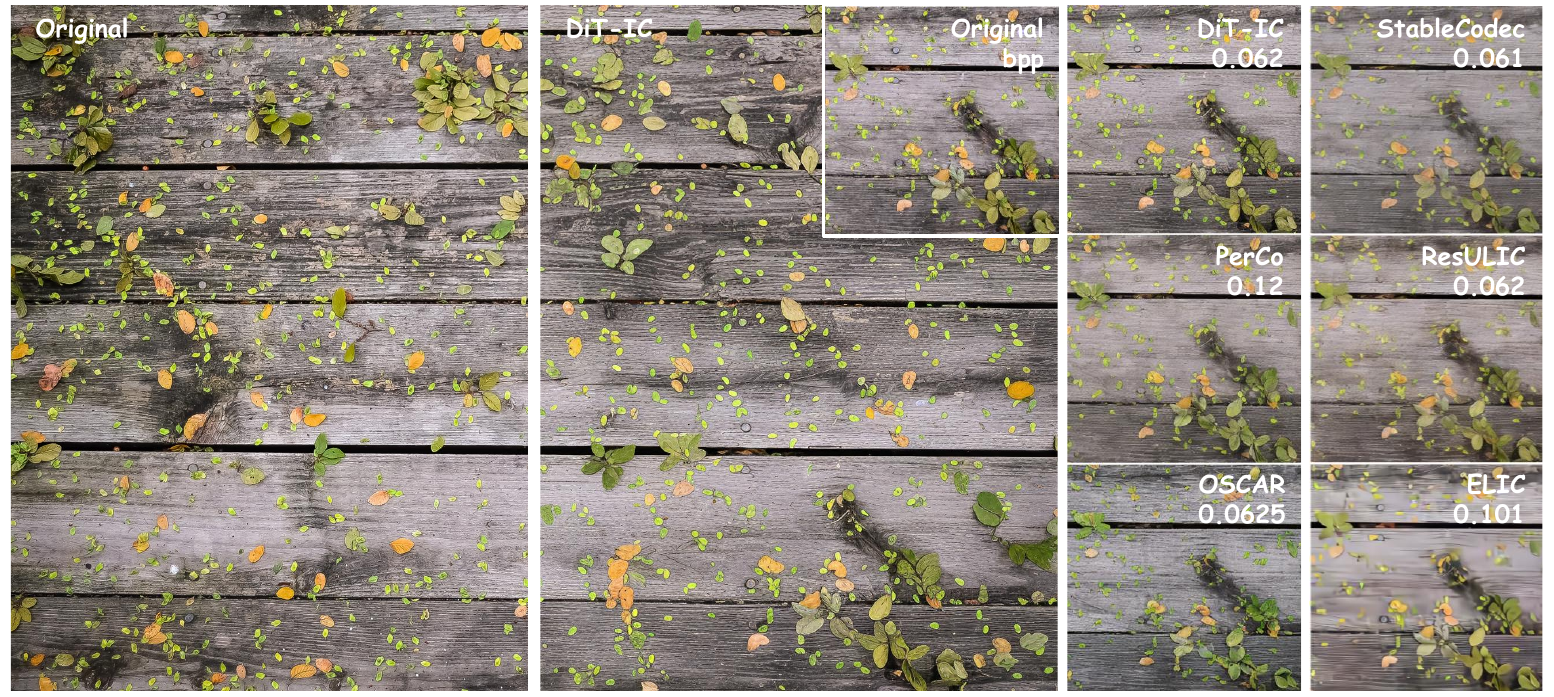}
    \caption{Visual examples and comparisons.}
    \label{fig:vis2}
\end{figure*}

\textbf{Perceptual Evaluation.}
To provide a comprehensive perceptual assessment beyond pixel-level measures, we additionally report several widely used perceptual metrics, including FID~\cite{heusel2017gans}, KID~\cite{binkowski2018demystifying}, NIQE~\cite{mittal2012making}, CLIPIQA~\cite{wang2023exploring} and MUSIQ~~\cite{kendall2017uncertainties}.
FID and KID measure the distributional discrepancy between reconstructed and reference images in the feature space of pretrained classifiers, serving as holistic indicators of realism. NIQE is a no-reference metric that evaluates natural scene statistics, reflecting perceived image naturalness. CLIPIQA leverages CLIP embeddings to assess semantic fidelity, while MUSIQ is modern deep IQA models designed to capture high-level perceptual quality across diverse content and resolutions.

As shown in Table~\ref{tab:result2}, we compare our DiT-IC with the state-of-the-art StableCodec ($\lambda=2.0$) at similar bitrates (approximately $0.03$--$0.04$ bpp). Due to differences in implementation environments, our reproduced results exhibit minor deviations from the originally reported values. We neglect the FID and KID results on Kodak as it is too small for calculating. DiT-IC achieves consistent improvements across most perceptual metrics on Kodak, DIV2K, and CLIC2020. The only exception is KID on CLIC2020, where StableCodec shows a slight advantage, but DiT-IC maintains overall superior perceptual performance across datasets and metrics..

Fig.~\ref{fig:fid} further presents FID as a function of bitrate. DiT-IC consistently outperforms prior codecs across operating points. The performance margin on CLIC is smaller than on DIV2K, likely because StableCodec is trained on CLIC, resulting in better dataset alignment. Similar trends are observed for KID. In addition, incorporating adversarial training further enhances perceptual realism, as evidenced by the comparison with \textit{Ours w/o Adv loss} in Fig.~\ref{fig:fid}.

\section{Visualization}
We provide additional qualitative results and comparisons on high-quality images from DIV2K~\cite{agustsson2017ntire} and CLIC 2020~\cite{toderici2020clic}. We compare our DiT-IC with representative compression models, including StableCodec~\cite{zhang2025stablecodec}, ELIC~\cite{he2022elic}, PerCo~\cite{korber2024perco}, OSCAR~\cite{guo2025oscar}, and ResULIC\cite{keultra}. As shown, DiT-IC delivers superior semantic consistency and textural realism while operating at lower bitrates than competing methods.

\begin{figure*}[ht]
    \centering
    \includegraphics[width=0.98\linewidth]{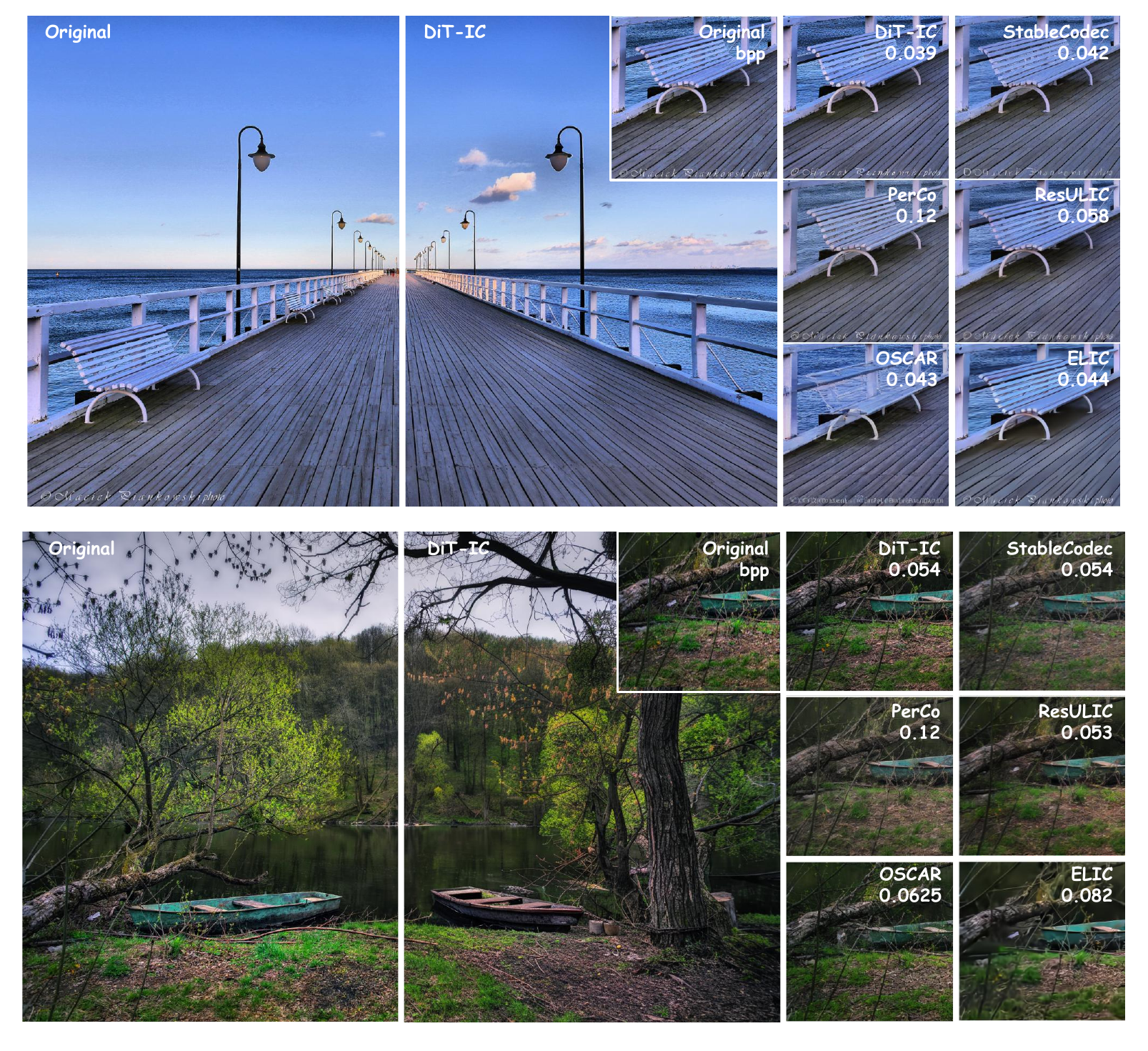}
    \caption{Visual examples and comparisons.}
    \label{fig:vis1}
\end{figure*}

\begin{figure*}[ht]
    \centering
    \includegraphics[width=0.96\linewidth]{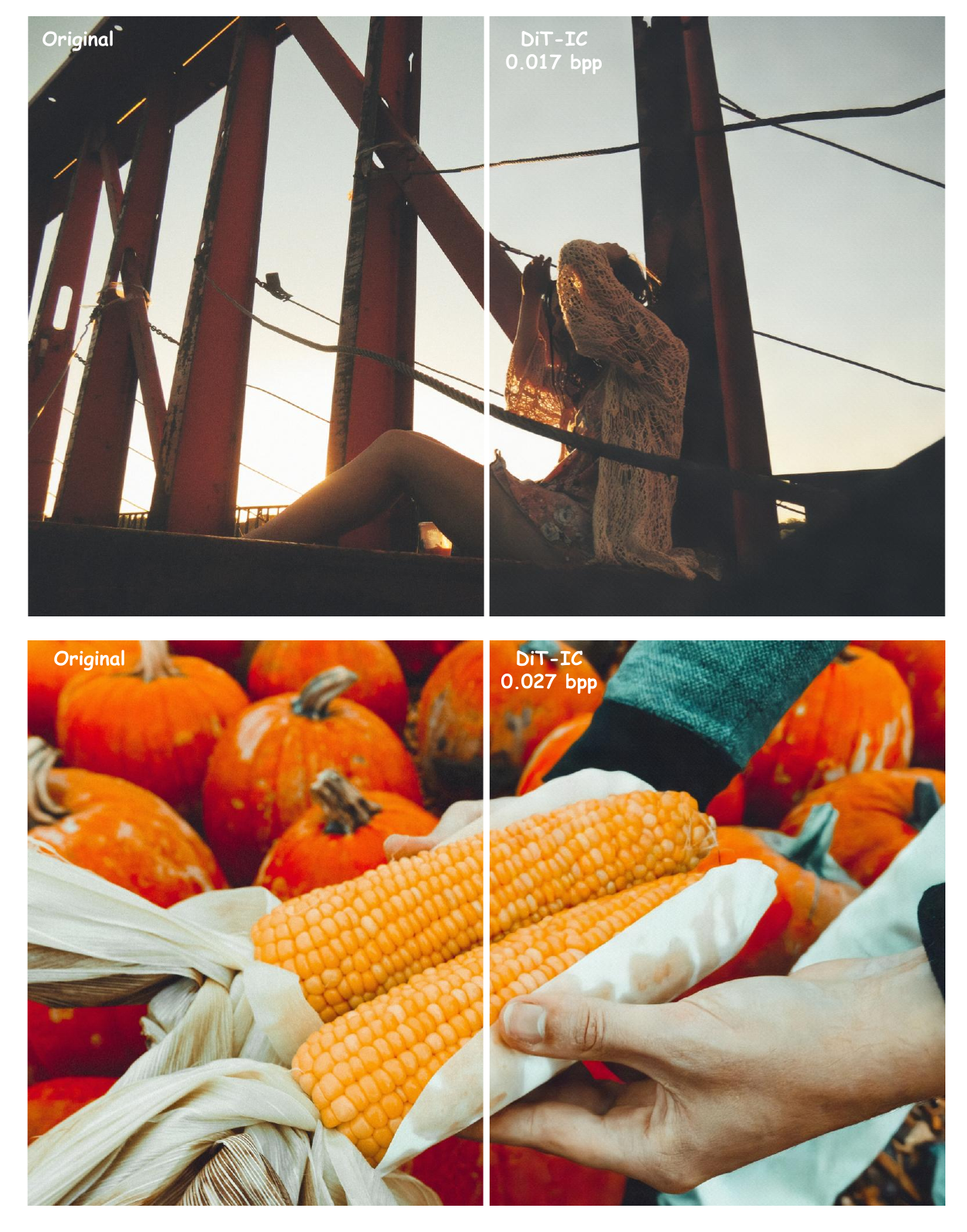}
    \caption{Visual examples and comparisons.}
    \label{fig:vis3}
\end{figure*}
\begin{figure*}[ht]
    \centering
    \includegraphics[width=0.95\linewidth]{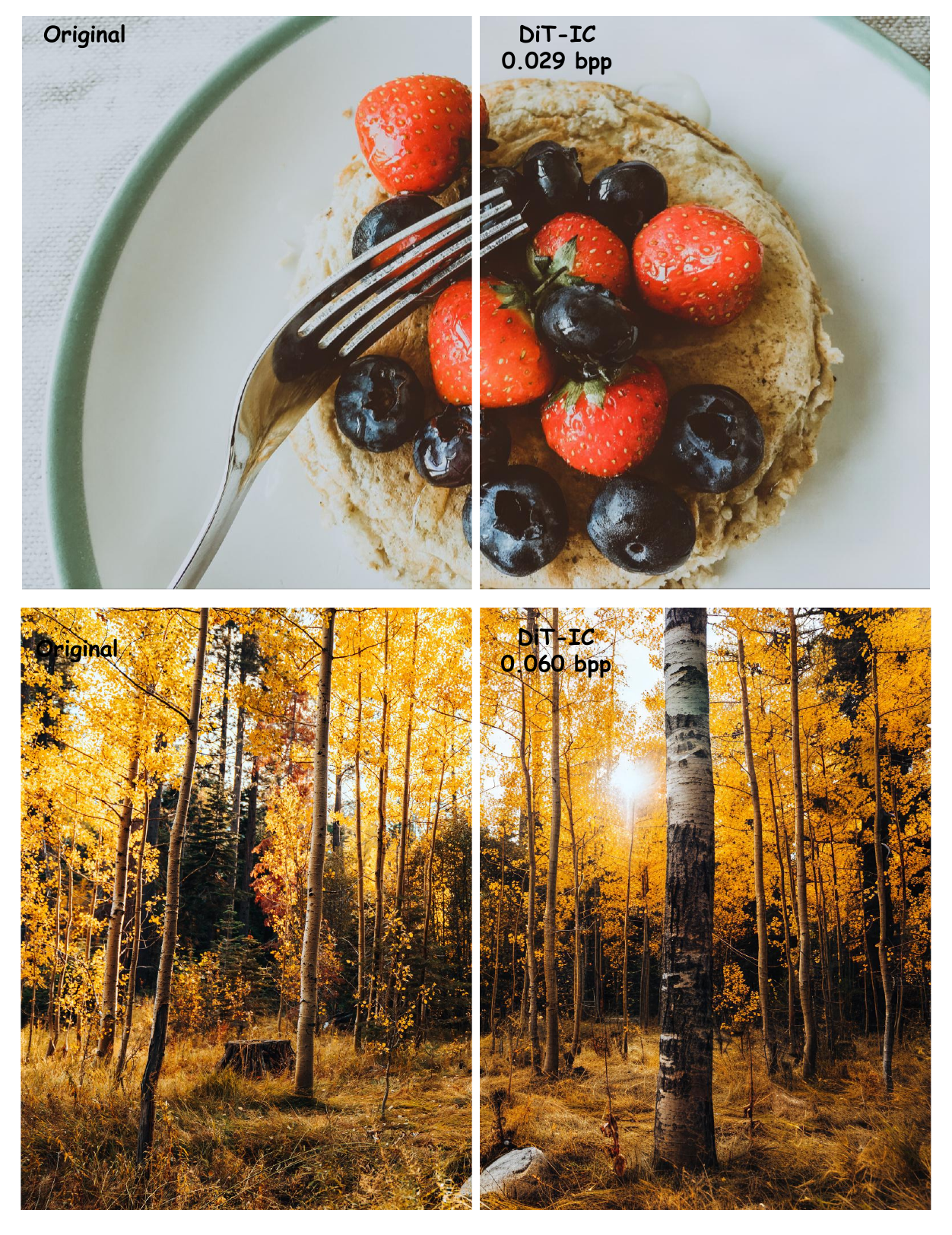}
    \caption{Visual examples and comparisons.}
    \label{fig:vis4}
\end{figure*}